%% file: main.tex
\documentclass{article}
\usepackage{graphicx}
\usepackage[margin=1in]{geometry}
\usepackage{amsmath, amssymb, amsthm, bbm, braket, complexity, mathtools}
\usepackage{xcolor}
\usepackage{hyperref}
\usepackage{cleveref}
\usepackage{parskip}
\usepackage{booktabs}
\usepackage{multirow}
\usepackage{caption}
\usepackage{subcaption}
\usepackage{authblk}
\usepackage{svg}
\usepackage{framed}
\usepackage[
maxnames=3,
minnames=3,
sorting=none,
style=numeric-comp, 
bibstyle=numeric-comp,
url=false,
isbn=false,
]{biblatex}
\RequirePackage[title,titletoc]{appendix}

\AtEveryBibitem{%
  \ifentrytype{article}{%
    \clearfield{publisher}%
    \clearfield{note}%
    \clearfield{editor}
    \clearfield{pages}
  }{}%
}

\usepackage{xfp}
\newcommand\SupplementaryMaterials{%
  \xdef\presupfigures{\arabic{figure}}
  \xdef\presupsections{\arabic{section}}
  \renewcommand\thefigure{S\fpeval{\arabic{figure}-\presupfigures}}
  \renewcommand\thesection{S\fpeval{\arabic{section}-\presupsections}}
}

\newenvironment{widetext}
    {
    }

\DeclarePairedDelimiter\pars{\lparen}{\rparen}

\newcommand{\bigo}[1]{\mathcal{O}\pars*{#1}}

\newtheorem{problem}{Problem}
\crefname{problem}{problem}{problems}

\title{Nonvariational quantum optimisation approaches to pangenome-guided sequence assembly}

\author[1, 2, 3]{Josh Cudby}
\author[2]{Sergii Strelchuk}

\date{\today}

\affil[1]{\small Department of Applied Mathematics and Theoretical Physics, University of Cambridge, Wilberforce Rd, Cambridge CB3 0WA, United Kingdom (jjcc2@cam.ac.uk)}
\affil[2]{Department of Computer Science, University of Oxford, Parks Rd, Oxford OX1 3QG, United Kingdom}
\affil[3]{Wellcome Sanger Institute, Hinxton, Cambridge CB10 1RQ, United Kingdom}

\begin{document}

\maketitle

\begin{abstract}

Assembling genomes from short-read sequencing data remains difficult in repetitive regions, where reference bias and combinatorial complexity limit existing methods. 
Pangenome-guided sequence assembly (PGSA) mitigates reference bias by reconstructing an individual genome as a walk through a population-level graph.
The associated problem, identifying a walk whose node visits match read-derived copy numbers, is NP-hard and already challenges classical solvers at a moderate scale. 
We develop near-term quantum optimisation approaches for this computational bottleneck. We consider two problem encodings: an established quadratic unconstrained binary optimisation and a new higher-order binary optimisation (HUBO) formulation.
The latter reduces the number of variables from $O(N^2)$ to $O(N\log N)$ and places moderate-sized instances within the qubit budget of current devices. 
We solve both using the Iterative-QAOA framework, which combines a fixed linear-ramp QAOA schedule with iterative warm-start bias updates, avoiding the overhead of full variational parameter optimisation. 
A custom circuit compilation strategy reduces hardware gate overhead by up to 67\% compared with standard tools.
In noiseless simulations of QUBO problems, Iterative-QAOA reliably identifies optimal assemblies from as few as $10^{-17}\%$ of all candidate solutions, and \textit{IBM} quantum hardware closely reproduces relevant results with sufficient sampling via CVaR-style post-selection. 
For HUBO, the variable reduction comes at the cost of deeper compiled circuits and greater noise sensitivity: an expected qubit--depth trade-off.
Our findings establish pangenome assembly as a concrete, biologically motivated problem class at the scale where quantum optimisation may first provide practical value.

\end{abstract}

\section{Introduction}
\input{input_files_v1/introduction}
\section{Results}
\label{sec:results}
\input{input_files_v1/results}
\section{Discussion}\label{sec:discussion}
\input{input_files_v1/discussion}

\section{Methods}
\label{sec:methods}
\input{input_files_v1/methods}

\section*{Acknowledgements}
The authors would like to thank the entire Quantum Pangenomics team for their valuable support and discussions throughout this work. 

\section*{Funding}
 Work on the Quantum Pangenomics project was supported by Wellcome Leap as part of the Q4Bio Program. 
 SS was also supported by the Royal Society University Research Fellowship.

\printbibliography

\pagebreak
\SupplementaryMaterials
\begin{center}
\textbf{\Large Supplementary Information}
\end{center}
\input{input_files_v1/appendix}

\end{document}

%% file: input_files_v1/introduction.tex


In~\cite{cudbyPangenomeguidedSequenceAssembly2026}, we introduced \textit{pangenome-guided sequence assembly} (PGSA), a method for assembling short reads from an individual using a high-quality precomputed pangenome as a reference.
We showed that PGSA attains assembly quality comparable to \textit{de novo }assemblers while producing substantially fewer contigs.
We proposed a \textit{Quadratic Unconstrained Binary Optimisation} (QUBO) formulation of PGSA, demonstrating improved robustness to noise arising from missing or divergent genomic regions compared to traditional search-based methods.
We also compared several classical optimisation strategies and provided a proof-of-principle for a quantum optimisation approach.
For a short genomics primer and background on sequence assembly, see~\cref{sec:bio_primer}.


In this work, we investigate the potential for quantum optimisation to provide practical utility for PGSA.
We develop a \textit{Higher-order Unconstrained Binary Optimisation} (HUBO) formulation of PGSA.
By encoding node indices in binary rather than enumerating node–time pairs, this formulation reduces the number of variables from $\bigo{TN}$ to $\bigo{T\log(N)}$.
We employ Iterative-QAOA (following~\cite{lopez-ruiz_non-variational_2025}) to solve both QUBO- and HUBO-encoded problems, exploring the trade-offs between quantum resources and solution quality.

\begin{figure}
    \centering
    \includegraphics[width=\textwidth]{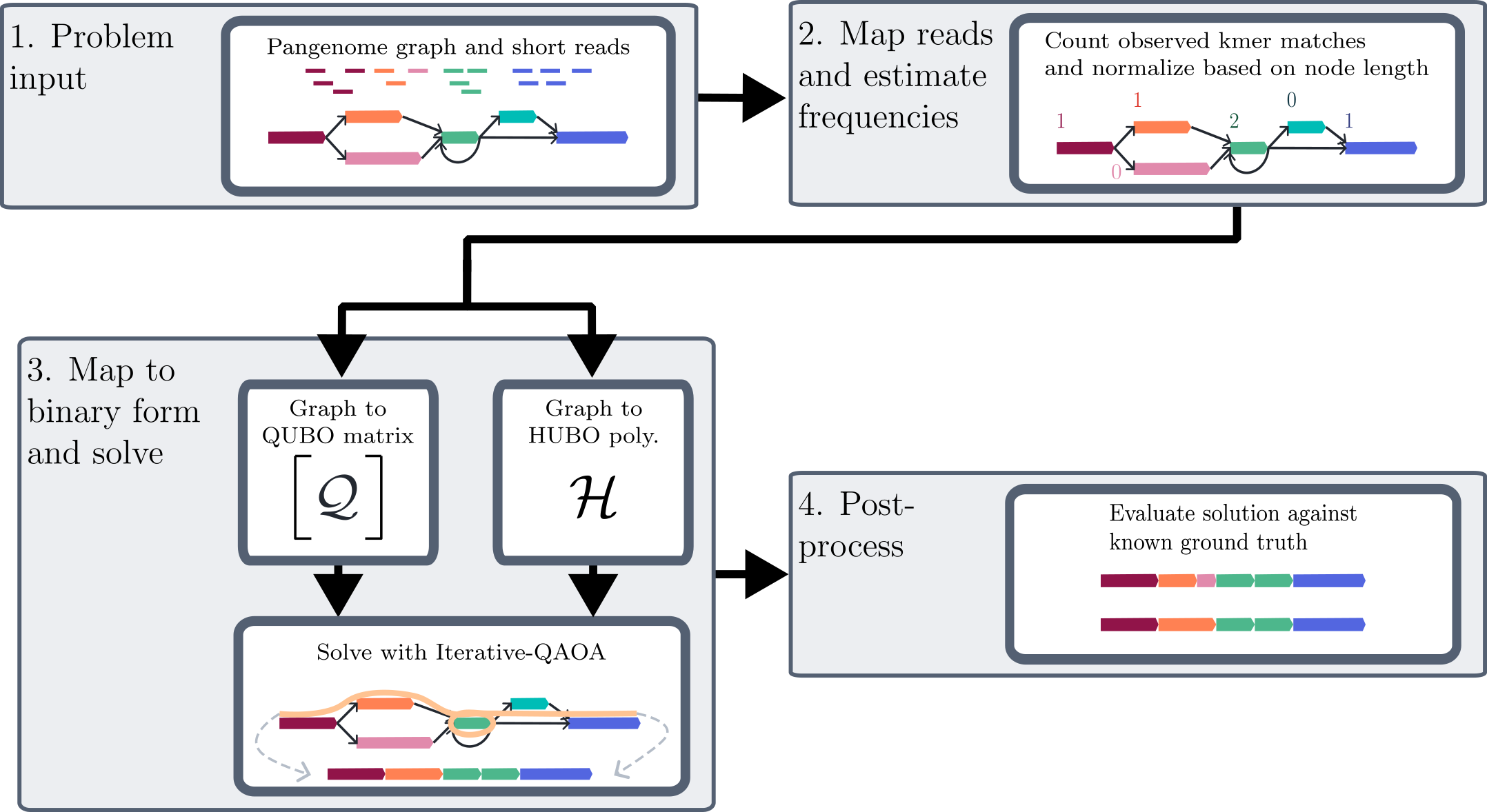}
    \caption{
        A sketch of the PGSA procedure. 
        (1) Problem inputs are a pangenome graph and a set of reads. 
        (2) Reads are matched onto nodes of the graph, and an estimate of the number of visits to each node is calculated. 
        (3) The path-finding problem is mapped to a binary optimization and solved with Iterative-QAOA.
        (4) Post-processing evaluates the quality of the assembly.
    }
    \label{fig:assembly_flowchart}
\end{figure}

PGSA comprises a multi-step workflow, shown in~\cref{fig:assembly_flowchart}, which can be summarised as:
\begin{description}
    \item[1. Problem inputs:] A synthetic pangenome and a randomly-sampled set of short fragments of DNA (or \textit{reads}) from a new genome.  
    \item[2. Map reads and estimate frequencies:] The reads are matched against the sequence data contained in the nodes of the graph. The number of appearances of each node in the new genome is estimated from the number of hits per node, normalised by the node length and the number of input reads.
    \item[3. Map to binary form and solve:] The goal of this step is to identify a path through the pangenome graph that best fits the observed frequencies, and it is the main computational bottleneck, referred to as \textit{Oriented Tangle Resolution}. We map to a binary optimization form and solve using Iterative-QAOA.
    \item[4. Post-process:] When the ground truth is known, as is the case for our synthetic problems in this work, we can evaluate the quality of the assembled genome compared.
\end{description}
For a full treatment, see~\cite{cudbyPangenomeguidedSequenceAssembly2026}.
In this work, we focus on the computational bottleneck in step three.

After step three, we have a pangenome graph $G$ whose vertex set $V$ is partitioned into two subsets $V_+$ and $V_-$, where a vertex $v_- \in V_-$ corresponds to the reverse complement of the DNA corresponding to $v_+ \in V_+$.
Each vertex has a corresponding weight, given by the function $w: V \rightarrow \mathbb{R}_+$.
This path-finding task is formalised as the following optimisation problem, which we call \emph{Oriented Tangle Resolution}.\\
\begin{problem}[Oriented Tangle Resolution]\label{prob:oriented_tangle}
    Given a vertex-weighted, oriented graph $G = (V_+\cup V_-, E, w)$, find a walk $W$ on $G$ that minimises 
    \begin{equation}\label{eq:oriented_tangle}
        C_G(W) = \sum_{v \in V} \Bigl( \#W(v_+) + \#W(v_-) - w(v) \Bigr)^2,
    \end{equation}
    where $\#W(v_+), \#W(v_-)$ are the number of times $W$ visits $v_+$ and $v_-$ respectively.
\end{problem}


Quantum optimisation offers a promising avenue for quantum utility and advantage, particularly for combinatorial problems in which the solution space grows exponentially.
QAOA~\cite{farhi_quantum_2014} addresses optimisation problems by mapping a cost function $C$ to a Hamiltonian $H_C$ whose ground state(s) correspond to optimal solution(s).
In its standard form, a QAOA circuit first prepares an initial state that is the ground state of a \textit{mixer} Hamiltonian, often choosing the simple $H_M \coloneqq - \sum_i X_i$ and corresponding initial state $\ket{+}^{\otimes n}$.
The state is then evolved by applying $p$ layers of alternating $\exp(-i\gamma_k H_C)$ and $\exp(-i\beta_k H_M)$ operations for real vectors $\beta =(\beta_1,\ldots,\beta_p)$ and $\gamma=(\gamma_1,\ldots,\gamma_p)$.
A classical optimiser is used to find optimal values of $\beta$ and $\gamma$ such that measurements of the final state are low-cost assignments with high probability. 

HUBO formulations allow higher-order interaction terms, enabling more compact encodings in this setting; the downside is that (classically) solving them is generally more difficult and less well-explored in the literature.
Standard classical approaches use a variable expansion to map HUBO instances onto QUBO instances.
Notably, however, HUBO may not have such a large solution overhead when mapped to QAOA algorithms on quantum computers.
In that setting, each layer of the cost Hamiltonian now contains multi-qubit $Z$ rotations, rather than only single- and two-qubit rotations.
These gates need to be compiled down to 2-qubit gates, but constructions that are linear in the number of qubits involved exist~\cite{glos_space-efficient_2020}.
Further depth reductions are possible through careful compilation (see~\cref{subsec:compilation}).


Due to their simple circuit structure, QAOA circuits are well-suited for implementation on Noisy Intermediate-Scale Quantum (NISQ) devices in the near term.
There is evidence that, for certain problem classes, QAOA offers an advantage over classical methods~\cite{shaydulin_evidence_2024,montanez-barrera_toward_2025,crooks_performance_2018}.
However, the potential utility of quantum optimisation is severely limited by hardware constraints.
The depth of QAOA circuits increases linearly with the hyperparameter $p$, and there is evidence that local quantum algorithms, such as $p=1$ QAOA, cannot offer any advantage over classical methods~\cite{barak_classical_2022}.
This creates a tension between increasing $p$ to enhance expressivity and keeping $p$ small enough for practical implementation.
Moreover, the variational loop involved in optimising the parameters is costly and can be challenging due to the presence of barren plateaus~\cite{fontana_characterizing_2024}.


Several strategies have been proposed to improve the performance of QAOA and to avoid the classical optimisation loop.
One approach is \textit{warm-starting} QAOA, where the initial state incorporates prior information rather than being uniform~\cite{egger_warm-starting_2021,WILLSCH2022108411,grant_initialization_2019}.
These strategies can improve convergence, avoid local minima and mitigate the effects of barren plateaus.
An orthogonal approach avoids the need for classical optimisation entirely by fixing variational parameters \textit{a priori}.
These parameters are chosen either by extrapolation from small problem instances~\cite{montanez-barrera_transfer_2025,shaydulin_parameter_2023,sureshbabu_parameter_2024} or from a parameter schedule that is known to work well in general~\cite{kremenetski_quantum_2023,montanez-barrera_toward_2025,dehn_extrapolation_2025,sakai_transferring_2025,PhysRevA.104.052419}.

\begin{figure*}[t]
    \centering
    \includegraphics[width=0.85\textwidth]{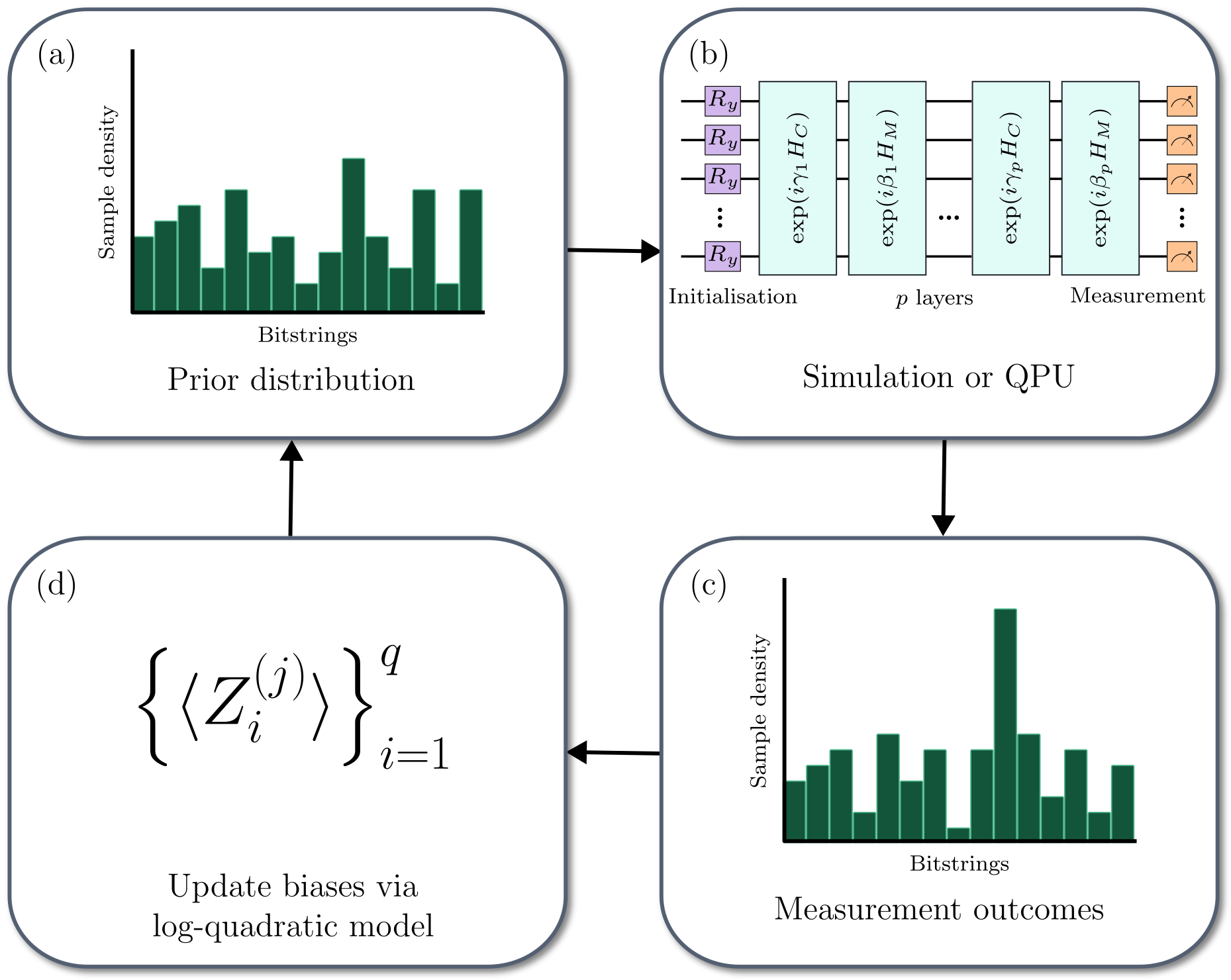}
    \caption{The workflow of the Iterative-QAOA algorithm. 
    (a) A prior distribution over bitstrings is initialised.
    (b) A non-variational ansatz is prepared by initialising qubits according to the prior distribution and then applying $p$ alternating layers of cost ($\exp(i\gamma H_C)$) and mixer ($\exp(i\beta H_M)$) unitaries. 
    (c) Samples are drawn from measurements on a physical quantum computer or from the final probability distribution in a classical simulation. 
    (d) The prior distribution is updated by computing new biases for each (qu)bit according to a log-quadratic model.
    }
    \label{fig:iterative_qaoa}
\end{figure*}

The authors of~\cite{lopez-ruiz_non-variational_2025} combine these approaches, leading to the Iterative-QAOA procedure.
In this framework, the bulk of the QAOA circuit is that of a fixed \textit{Linear Ramp} QAOA (LR-QAOA)~\cite{kremenetski_quantum_2023}, where the parameters follow a linear schedule:
\begin{equation}
    \beta_k \coloneqq  \left(1 - \frac{2k-1}{2p}\right) \Delta_\beta, \qquad \gamma_k \coloneqq  \frac{2k-1}{2p} \Delta_\gamma.
\end{equation}
Here, $\Delta_\beta$ and $\Delta_\gamma$ are hyperparameters that control the ramp ranges.
To improve the performance of LR-QAOA, Iterative-QAOA refines the warm-started initial state after each run for several iterations~\cite{yuan_iterative_2025}, using the measured bitstrings $s_j$ and corresponding energies $E_j \coloneqq \braket{s_j | H_C | s_j}$ to compute initial biases for each qubit at the next step.
A summary of the method is provided in~\cref{subsec:iterative_qaoa} and a schematic is shown in~\cref{fig:iterative_qaoa}.
This method progressively guides the search towards regions of Hilbert space with low energies while requiring far fewer iterations than a full variational optimisation.

We view both the QUBO and HUBO encodings as complementary steps toward practical quantum approaches for PGSA.
For a given qubit count, QUBO circuits are shallower and are therefore easier to implement on current hardware; we demonstrate feasibility on available devices.
HUBO circuits present greater implementation challenges on current machines, but they can encode problems using fewer binary variables and thus fewer qubits.
Current devices, such as \textit{IBM Boston}, provide sufficient qubits to encode graph instances of up to 64 nodes for walks of length up to 26.
Problems of this scale can already be challenging for classical solvers, suggesting a plausible path toward quantum utility as hardware fidelity and optimisation methods improve.

Maurizio and Mazzola~\cite{maurizioQuantumComputingGenomics2025} recently surveyed quantum computing for genomics and argued that near-term quantum speedups are restricted to problems whose core optimisation is hard for classical approximate solvers and require a limited number of variables. 
Our Oriented Tangle Resolution satisfies both criteria. The problem generalises Hamiltonian path finding and is NP-hard in general. 
The existing state-of-the-art prior to~\cite{cudbyPangenomeguidedSequenceAssembly2026} \textit{pathfinder} proceeds via exhaustive search, demonstrating that this is a hard problem with no known approximation schemes.
In~\cite{cudbyPangenomeguidedSequenceAssembly2026}, the classical QUBO solvers \textit{Gurobi} (branch-and-bound) and \textit{MQLib} were competitive but required multiple restarts and produced variable solution quality across instances, with neither consistently dominating. 
The solution space grows exponentially with walk length $T \sim N$, and the QUBO encoding requires $\bigo{N^2}$ variables. 
A novel HUBO encoding introduced here reduces this to $\bigo{N\log(N)}$, placing moderate-size instances within the qubit budget of current and near-term devices.

The remainder of the paper is organised as follows.
In~\cref{sec:results} we present numerical and hardware results; \cref{sec:discussion} summarises the findings and discusses future directions.
\Cref{sec:methods} provides full method details, including cost function construction, circuit compilation and Iterative-QAOA hyperparameters.

%% file: input_files_v1/results.tex
\subsection{Experimental Setup}
All problem instances were generated from small synthetic pangenome graphs.
Circuit generation and transpilation were handled by our hand-rolled transpiler (see~\cref{subsec:compilation}).

Simulations were performed using noiseless matrix-product state (MPS) simulations with a fixed bond dimension.

Hardware experiments were executed on \textit{IBM Boston} with Pauli Twirling~\cite{wallman_noise_2016} enabled.
We used a Conditional Value at Risk (CVaR) loss function, effectively oversampling and keeping only a fraction $\alpha$ of the highest-performing samples (see~\cref{subsec:iterative_qaoa}).

We set the hyperparameters $\Delta_\beta $ and $\Delta_\gamma $ according to the results of the parameter exploration, detailed in~\cref{sec:param_exploration}.
For QUBO instances, we took $\Delta_\beta = 0.63$ and $\Delta_\gamma = 0.16$, and for HUBO instances, we used $\Delta_\beta = 0.75$ and $\Delta_\gamma = 0.30$.
These values were chosen to perform best for $p=1$ circuits, which are the most realistic to access on current-term quantum devices.

\subsection{QUBO Problems}


We first benchmarked Iterative-QAOA on QUBO instances up to 80 qubits and beyond.
We tested with 1, 3 and 5 layers of the LR-QAOA ansatz\"e for 5 iterations, simulating 40,000 samples per iteration.

\begin{figure}
    \centering
\includegraphics[width=0.85\textwidth]{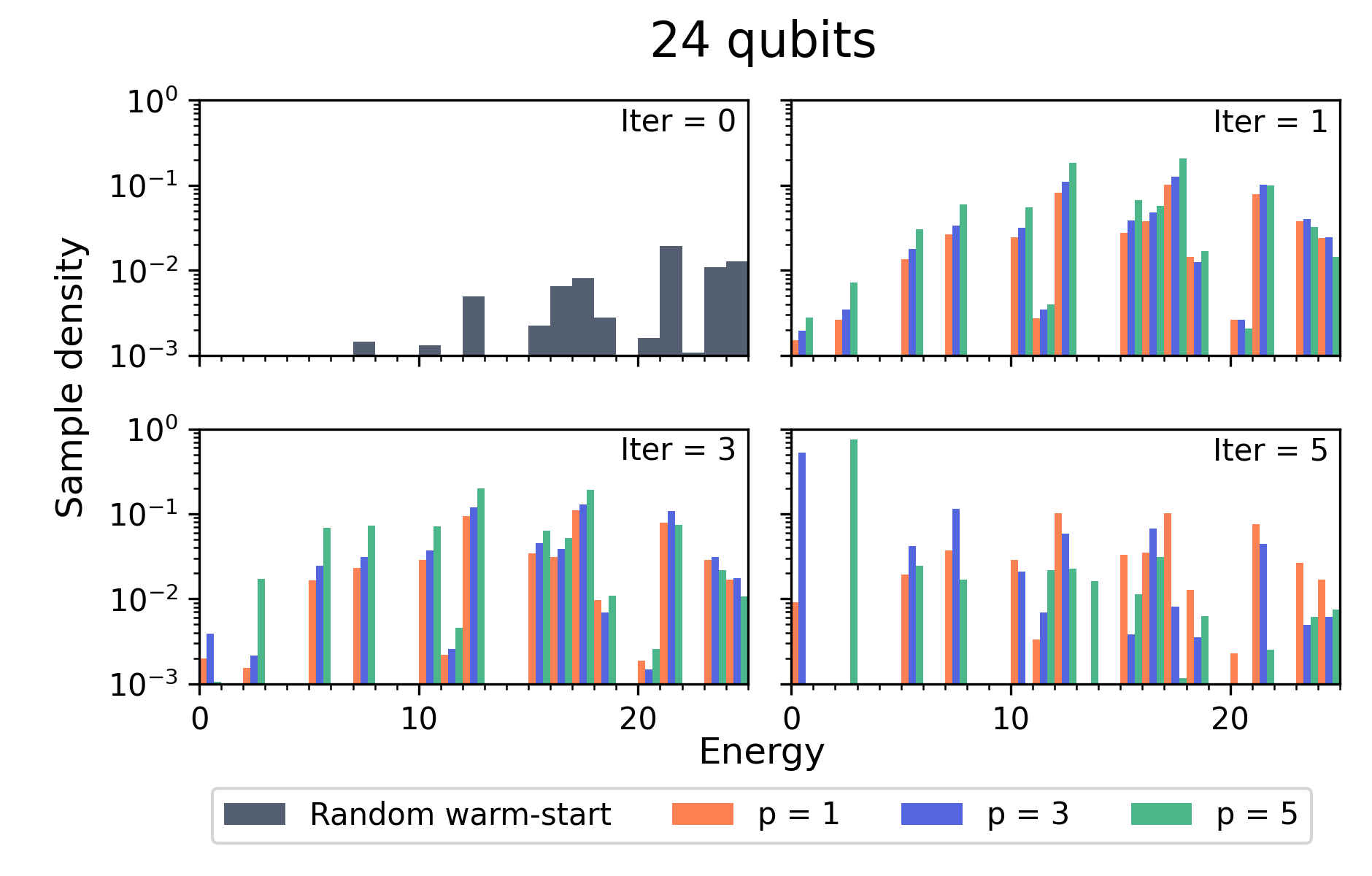}
    \caption{
        Noise-free MPS simulation of Iterative-QAOA performance on a representative 24-qubit QUBO instance. 
        Energy distribution over iterations $0$ to $5$ for LR-QAOA depths $p =  1,\,3,\,5$  with 40,000 samples per iteration.
        ($\Delta_\beta=0.63, \ \Delta_\gamma=0.16.$)
    }
    \label{fig:qubo_sim_24_qubits}
\end{figure}

\begin{figure}
    \centering
\includegraphics[width=0.85\textwidth]{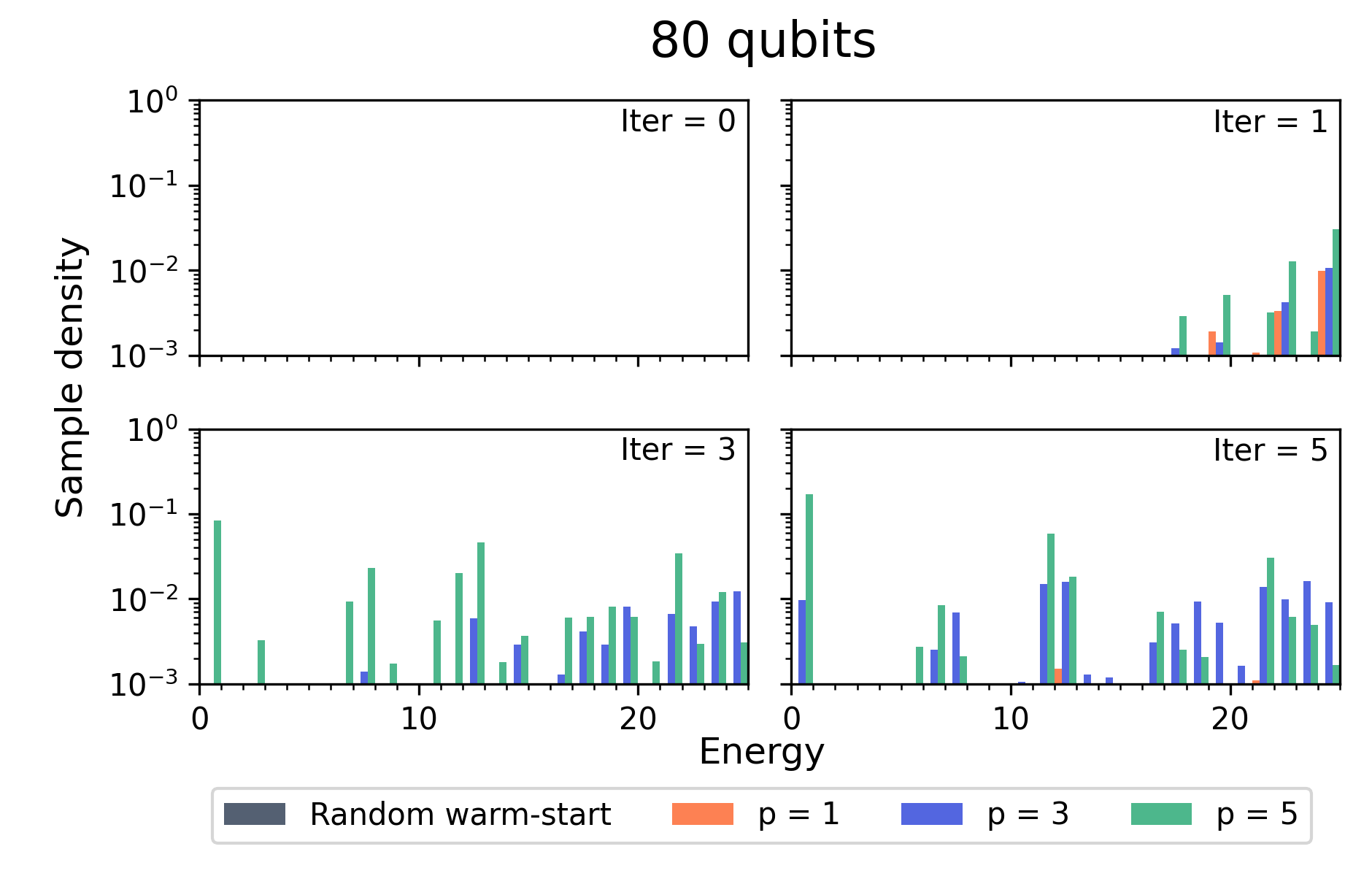}
    \caption{
        Noise-free MPS simulation of Iterative-QAOA performance on a representative 80-qubit QUBO instance. 
        Energy distribution over iterations $0$ to $5$ for LR-QAOA depths $p =  1,\,3,\,5$  with 40,000 samples per iteration.
        ($\Delta_\beta=0.63, \ \Delta_\gamma=0.16.$)
    }
    \label{fig:qubo_sim_80_qubits}
\end{figure}

\Cref{fig:qubo_sim_24_qubits} shows a representative 24-qubit instance, and \cref{fig:qubo_sim_80_qubits} shows the corresponding results for an 80-qubit instance.

For the 24-qubit instance, the original prior distribution yields some low-energy solutions but no optimal ones. After a single LR-QAOA run, optimal solutions are sampled for all values of $p$; in practice, this would be sufficient to terminate the algorithm early.
Over subsequent iterations, we observe that the $p=1$ circuit gradually concentrates probability mass on the optimal solutions, whereas the $p=3$ circuit rapidly enters a regime in which more than $50\%$ of samples are optimal.
Conversely, the $p=5$ circuit converges strongly to a slightly suboptimal solution with energy $2$.
This behaviour likely reflects a choice of schedule parameters that is not well matched to this instance at $p=5$.
Nonetheless, sampling the optimum even once is sufficient to solve the underlying problem, so we count the run as a success.

For the 80-qubit instance, the prior distribution produces no low-energy samples.
After four iterations (not shown), the $p=3$ circuit samples optimal solutions, whereas the $p=5$ circuit requires only three iterations.
By contrast, the $p=1$ circuit drifts towards higher energies, indicating insufficient circuit expressivity at this problem size.
Even using all five iterations of 40,000 samples corresponds to exploring only a $1.65\times 10^{-19}$ fraction of the search space, highlighting the efficiency of Iterative-QAOA.

We then performed hardware experiments up to 48 qubits on the \textit{IBM Boston} Heron R3 device.
We choose the total number of samples so that the updated biases are computed from 4,000 ``good'' samples.
We use $p=1$ circuits to minimise depth-induced noise, observing that they are sufficiently expressive to solve all problem instances in simulation except the 80-qubit instance.

\begin{figure}
    \centering
\includegraphics[width=0.85\textwidth]{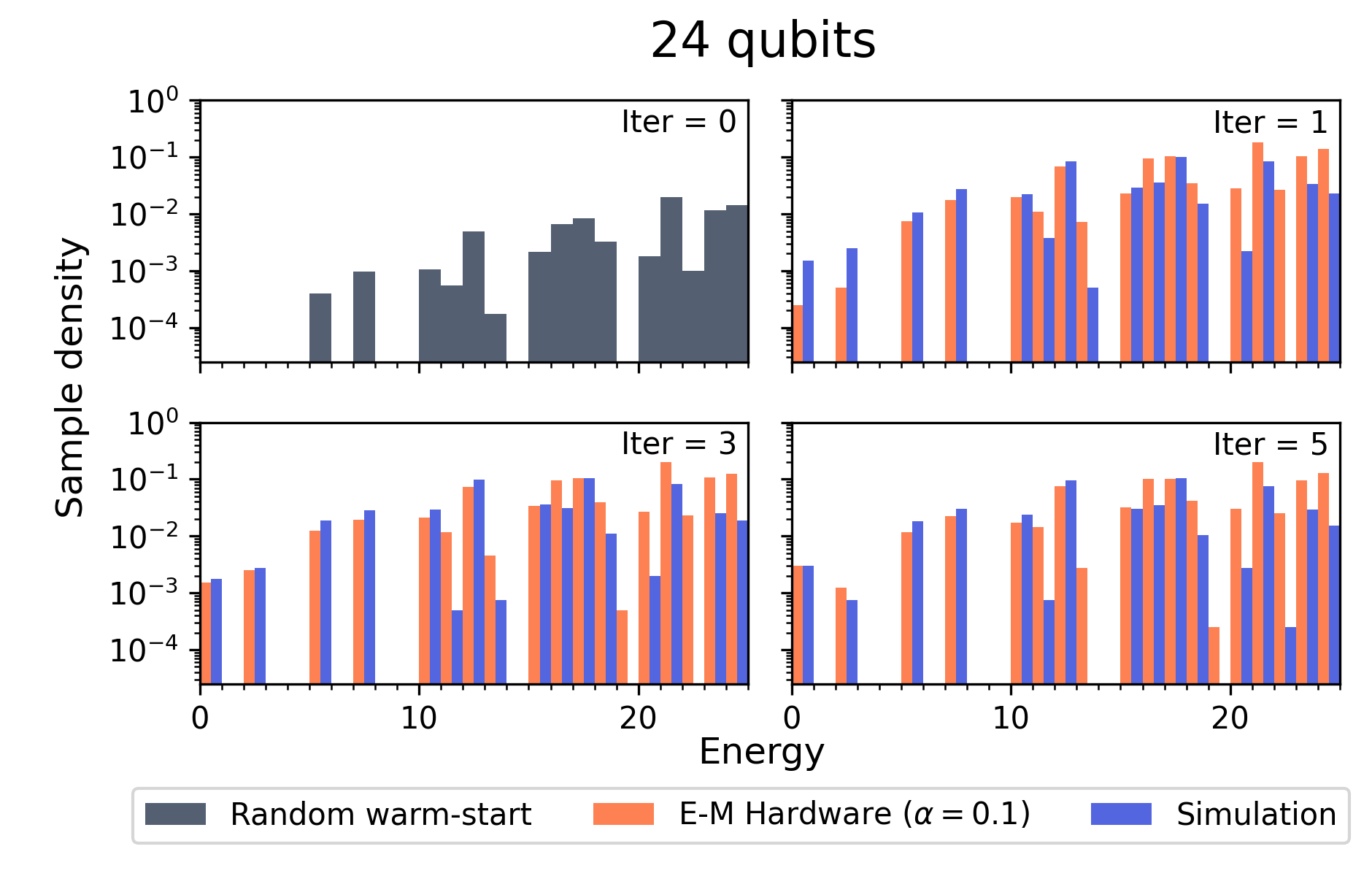}
    \caption{
        Iterative-QAOA on quantum hardware for a 24-qubit QUBO instance with $p=1$ and CVaR-style post-selection: comparison between error-mitigated hardware (best 4,000 shots per iteration from 40,000 total, \textit{i.e.}, $\alpha=0.1$) and a 4,000-samples-per-iteration noise-free simulation. 
        ($\Delta_\beta=0.63, \ \Delta_\gamma=0.16.$)
    }
    \label{fig:qubo_hardware_24_qubits}
\end{figure}

For the moderately sized 24-qubit instance, we found that $\alpha = 0.1$ (40,000 samples per iteration) was sufficient to accurately reflect the noise-free simulation results.
This circuit contained a total of 3,423 operations, including 670 error-dominating $CZ$ gates.
\Cref{fig:qubo_hardware_24_qubits} compares the error-mitigated hardware results (\textit{i.e.}, the best 4,000 shots per iteration) with a noise-free simulation using 4,000 samples per iteration.
We see that both the simulation and the experiment sample the optimum in the first iteration and follow similar convergence trajectories.
This example highlights the power of the CVaR method when samples are cheap, allowing us to harness the full potential of the Iterative-QAOA method.

\begin{figure}
    \centering
\includegraphics[width=0.85\textwidth]{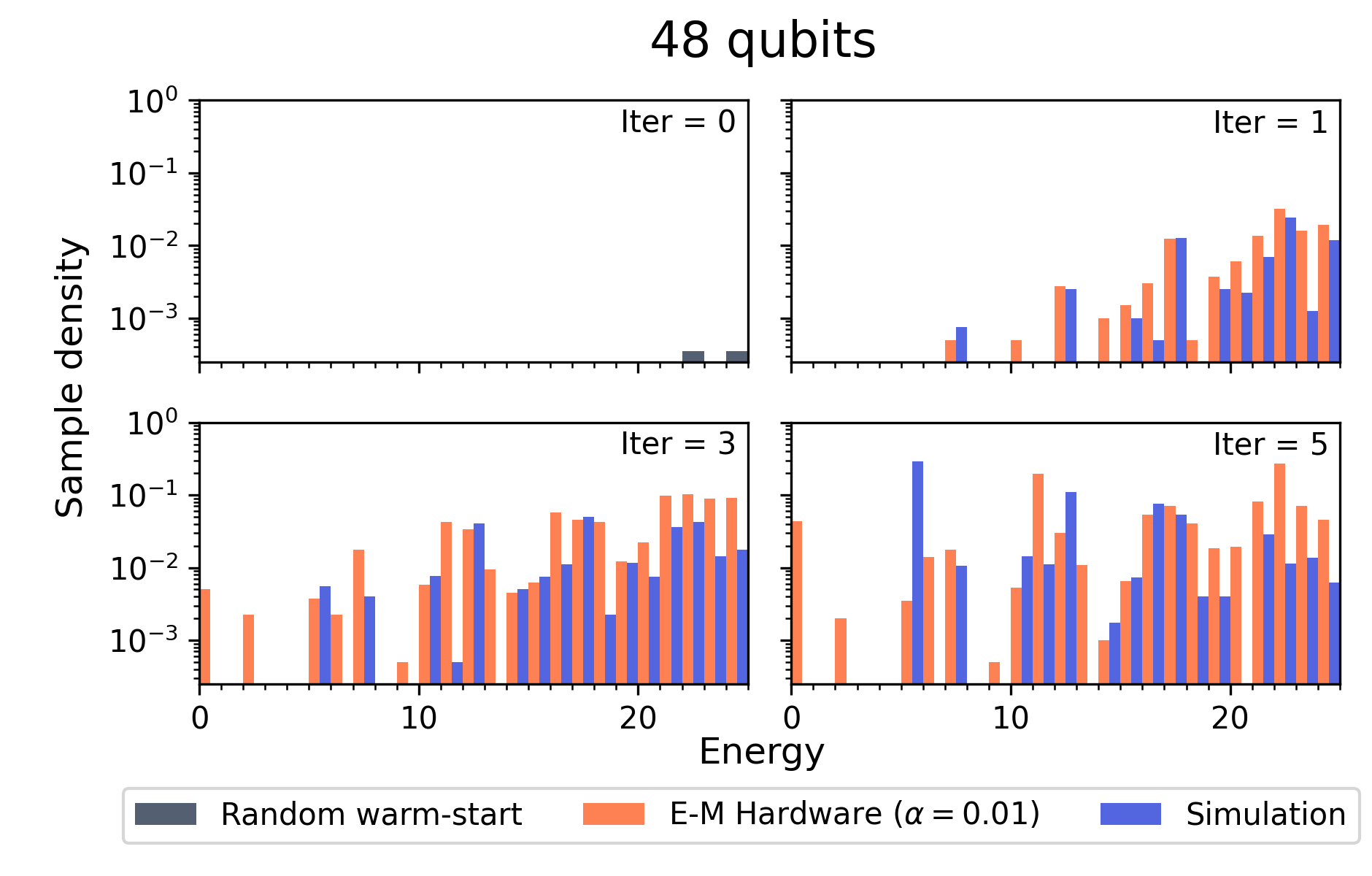}
    \caption{
            Iterative-QAOA on quantum hardware for a 48-qubit QUBO instance with $p=1$ and CVaR-style post-selection: comparison between error-mitigated hardware (best 4,000 shots per iteration from 400,000 total, \textit{i.e.}, $\alpha=0.01$) and a 4,000-samples-per-iteration noise-free simulation. 
            An optimal solution is observed at iteration 3 for the hardware experiment.
            ($\Delta_\beta=0.63, \ \Delta_\gamma=0.16.$)
    }
    \label{fig:qubo_hardware_48_qubits}
\end{figure}

For the larger 48-qubit instance, we decreased $\alpha$ to $0.01$ to improve performance, requiring 400,000 samples per iteration.
This circuit had $13{,}273$ operations with $2{,}865$  2-qubit gates.
The results are shown in~\cref{fig:qubo_hardware_48_qubits}.
With these settings, the experiment outperforms the simulation due to the breadth of available samples.
Notably, the experiment sampled the optimum during iteration three and has largely converged there by iteration five, whereas the simulation converges strongly to an energy-$5$ solution.

\subsection{HUBO Problems}

\begin{figure}
    \centering
\includegraphics[width=0.85\textwidth]{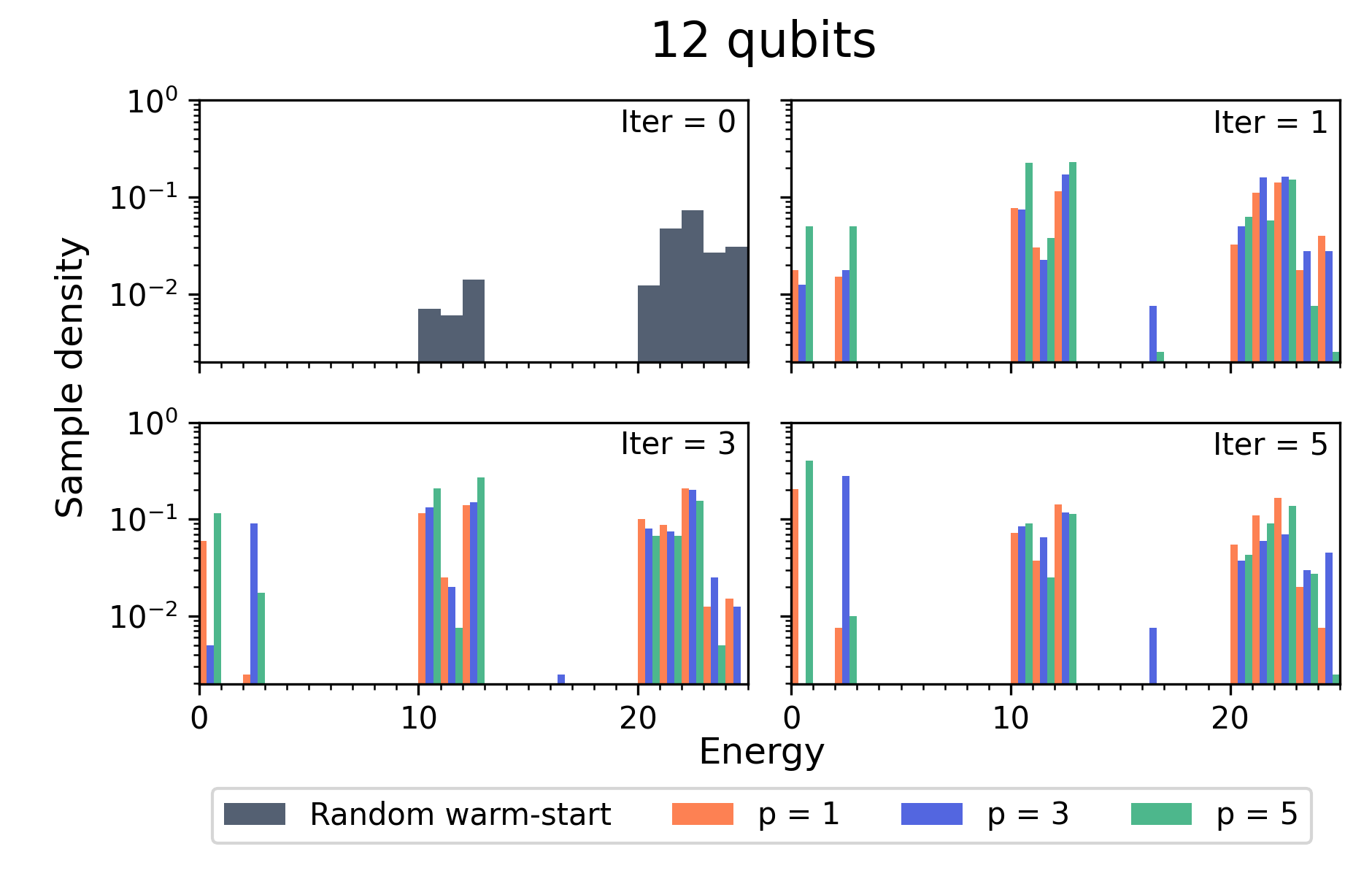}
    \caption{
        Noise-free MPS simulation of Iterative-QAOA performance on a representative 12-qubit HUBO instance. 
        Energy distribution over iterations $0$ to $5$ for LR-QAOA depths $p =  1,\,3,\,5$  with 400 samples per iteration.
        ($\Delta_\beta=0.75, \ \Delta_\gamma=0.30.$)
    }
    \label{fig:hubo_sim_12_qubits}
\end{figure}

\begin{figure}
    \centering
\includegraphics[width=0.85\textwidth]{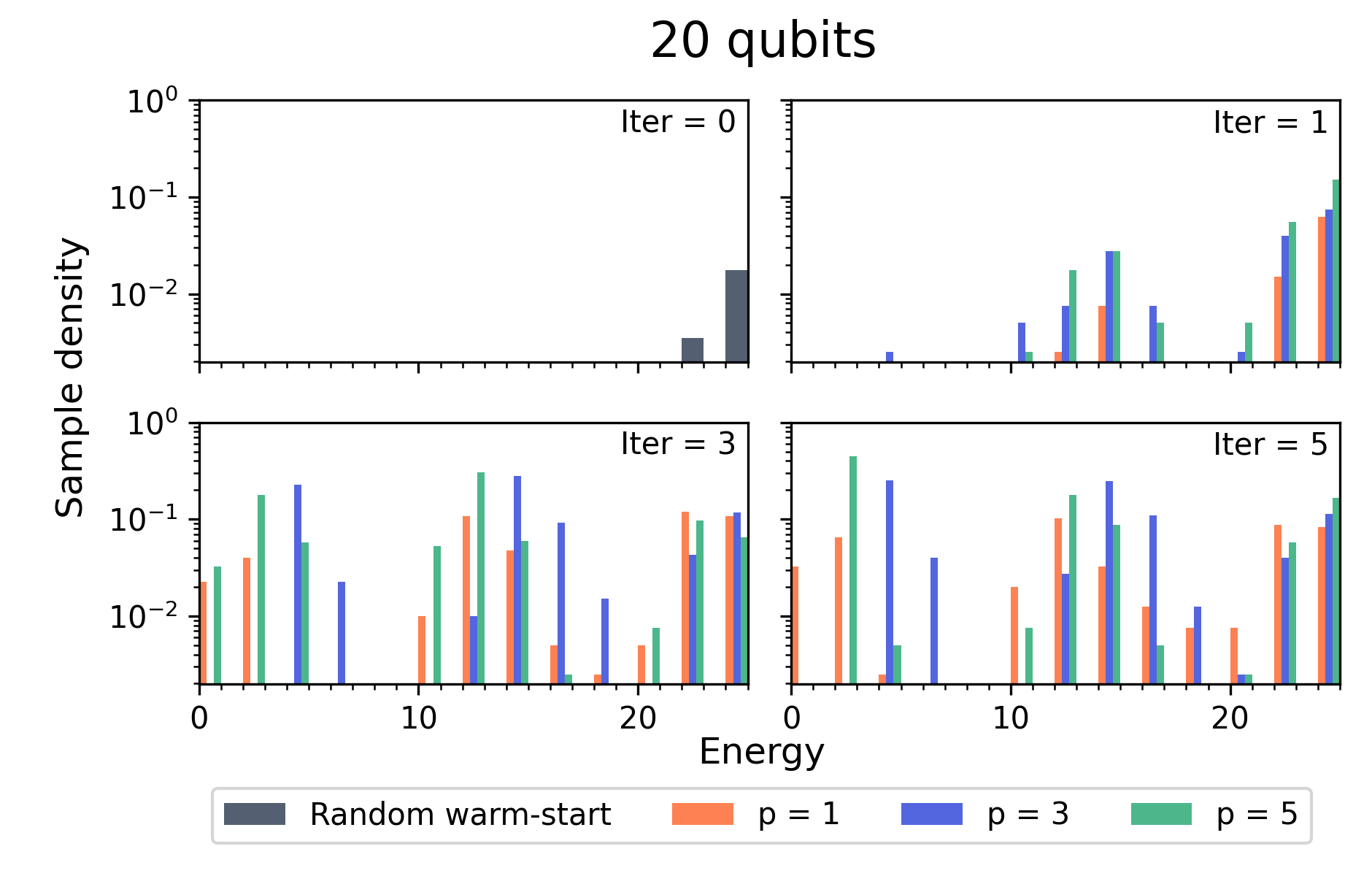}
    \caption{
        Noise-free MPS simulation of Iterative-QAOA performance on a representative 20-qubit HUBO instance. 
        Energy distribution over iterations $0$ to $5$ for LR-QAOA depths $p =  1,\,3,\,5$  with 400 samples per iteration.
        ($\Delta_\beta=0.75, \ \Delta_\gamma=0.30.$)
    }
    \label{fig:hubo_sim_20_qubits}
\end{figure}

We next consider the HUBO formulation, which reduces qubit count at the expense of higher-order interactions.
Since the circuits, for a fixed number of qubits, are deeper than the QUBO versions, we focus on smaller problem instances.
Since the problems are smaller, we take fewer shots to artificially increase their difficulty, with only 400 shots per iteration in the simulations. 
Results for 12 and 20 qubits are shown in~\cref{fig:hubo_sim_12_qubits,fig:hubo_sim_20_qubits}.

For the 12-qubit instance, all circuit depths sampled the optimal several times during the first iteration, whereas the unbiased random initialisation has very low density on optimal solutions.
In practice, all of the algorithms would therefore terminate after only 400 samples, less than 10\% of the sample space.
Interestingly, the $p=3$ circuit converges to a non-optimal solution, again highlighting the non-trivial relationship among parameter choice, circuit depth, and the problem instance; this behaviour may also be an artefact of sample variance due to the very few shots taken.


For the 20-qubit instance, this behaviour is exaggerated.
The $p=1$ circuit successfully converges a moderate amount of probability mass to the optimal.
Although the $p=5$ circuit samples the optimal in iteration 3, it eventually converges to non-optimal solutions.
The $p=3$ circuit fails to sample the optimal at all.


We also ran these circuits on the \textit{IBM Boston} device, again choosing $p=1$ to minimise the effects of noise.
We chose $\alpha$ so that 400 samples were used for the bias updates at each step.

\begin{figure}
    \centering
\includegraphics[width=0.85\textwidth]{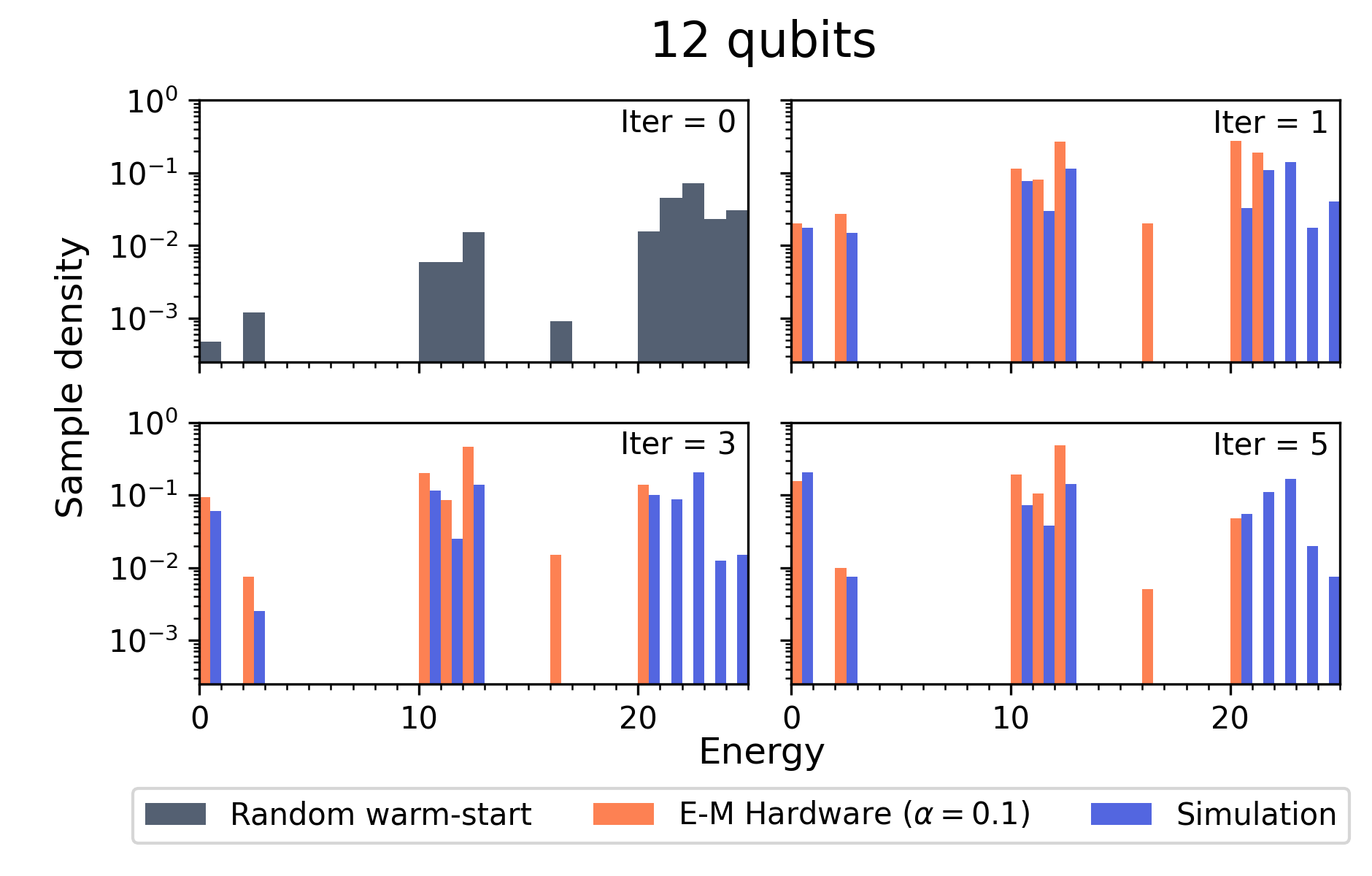}
    \caption{
        Iterative-QAOA on quantum hardware for a 12-qubit HUBO instance with $p=1$ and CVaR-style post-selection: comparison between error-mitigated hardware (best 400 shots per iteration from 4,000 total, \textit{i.e.}, $\alpha=0.1$) and a 400-samples-per-iteration noise-free simulation. 
        ($\Delta_\beta=0.75, \ \Delta_\gamma=0.3.$)
    }
    \label{fig:hubo_hardware_12_qubits}
\end{figure}

For the smaller 12-qubit instance, we took $\alpha = 0.1$ for a total of 4,000 samples per iteration; this circuit has 3,024 operations and 697 error-dominating $CZ$ gates.
Results are shown in~\cref{fig:hubo_hardware_12_qubits}, and we see good agreement between the noiseless simulation and the error-mitigated hardware results.

\begin{figure}
    \centering
\includegraphics[width=0.85\textwidth]{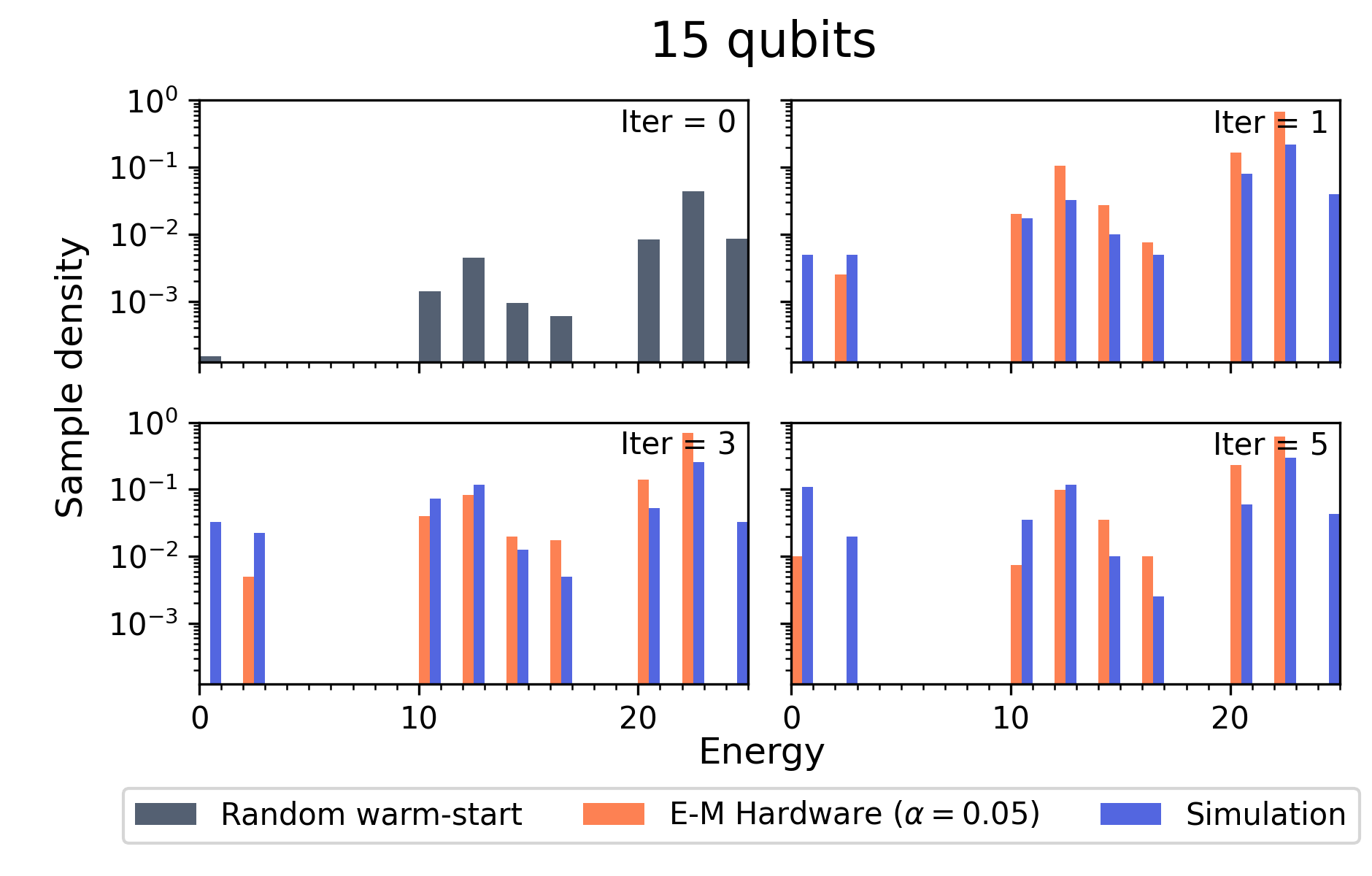}
    \caption{
        Iterative-QAOA on quantum hardware for a 15-qubit HUBO instance with $p=1$ and CVaR-style post-selection: comparison between error-mitigated hardware (best 400 shots per iteration from 8,000 total, \textit{i.e.}, $\alpha=0.05$) and a 400-samples-per-iteration noise-free simulation. 
        ($\Delta_\beta=0.75, \ \Delta_\gamma=0.3.$)
    }
    \label{fig:hubo_hardware_15_qubits}
\end{figure}

For the larger 15-qubit instance, we decreased $\alpha$ to 0.05, leading to 8,000 samples per iteration. 
This circuit has 5,124 operations, of which 1,219 are two-qubit $CZ$ gates.
Results are shown in~\cref{fig:hubo_hardware_15_qubits}. The optimal was first sampled in iteration 4 (not shown), whereas the simulation starts sampling the optimum during the first iteration.
With the high number of two-qubit gates, more aggressive error mitigation strategies are needed to recover the performance of classical noiseless simulations.

%% file: input_files_v1/discussion.tex
This work investigates the potential for near-term quantum optimisation to address a computational bottleneck in PGSA: identifying a walk through a pangenome graph whose node visitation frequencies match copy-number estimates derived from read mapping. 
We studied two binary optimisation encodings of the resulting Oriented Tangle Resolution objective: an established QUBO formulation, and a new HUBO formulation. 
Both were benchmarked using the Iterative-QAOA framework in noiseless simulation and on IBM quantum hardware.

\subsection{Summary of main findings}
For QUBO instances, Iterative-QAOA reliably identifies optimal solutions in simulation at moderate circuit depths. The bias-updating mechanism concentrates probability mass on low-energy states within a small number of iterations. On a representative 24-qubit instance, optimal solutions appear after a single iteration even at shallow depth, and increasing depth accelerates probability concentration.

At larger scales (e.g., 80 qubits), we observe a qualitative separation in behaviour: intermediate depths ($p=3$ and $p=5$) reach optimal solutions within a few iterations, whereas very shallow circuits ($p=1$) can stagnate or drift toward higher-energy states. This indicates that larger instances require additional circuit depth to represent correlations induced by the QUBO constraints.

Hardware experiments highlight a complementary trade-off.
Although deeper circuits perform well in simulation, current devices favour small $p$ to limit noise accumulation. 
Using $p=1$ and CVaR-style post-selection, we obtain trajectories that qualitatively match noiseless simulation on a 24-qubit instance.
On a 48-qubit instance, optimal solutions are observed when enough total shots are available. 
In the current depth-limited regime, performance depends not only on circuit expressivity but also on the sampling budget and the use of post-selection.

For HUBO, our results show both the promise and the challenges of higher-order formulations in a near-term setting. 
The HUBO formulation reduces variable count by encoding node indices in binary but introduces multi-qubit interactions in the cost Hamiltonian, leading to deeper compiled circuits.
In noiseless simulations of smaller instances, Iterative-QAOA samples optimal solutions efficiently.
However, behaviour with increasing depth is not always monotonic: in some cases, deeper circuits transiently sample optimal states before concentrating on suboptimal modes. 
Because HUBO experiments used relatively few shots per iteration, this effect may partially reflect shot noise interacting with the bias update rule; moreover, hyperparamters were chosen to favour the small $p$ circuits.
Nevertheless, it demonstrates that increased depth does not guarantee improved performance without schedule calibration.

On hardware, HUBO experiments further illustrate the depth–noise trade-off. For smaller instances, error-mitigated results align with noiseless simulation under CVaR post-selection. For larger instances with higher two-qubit gate counts, optimal solutions appear only at later iterations, and deviations from simulation increase. This suggests that improved error mitigation and compilation will be necessary as interaction order and two-qubit density grow.

\subsection{Choosing QUBO vs.\ HUBO in the QAOA setting}
The primary motivation for HUBO is qubit efficiency: reducing the number of binary variables lowers qubit requirements, which is valuable when quantum memory is the limiting resource.
However, this reduction increases circuit depth because higher-order terms must be implemented as multi-qubit $Z$ rotations and decomposed into two-qubit gates. 
Our results reflect this trade-off. 
QUBO instances scale to larger qubit counts in simulation and admit relatively shallow cost layers, whereas HUBO instances require smaller problem sizes (for a fixed depth budget) and become more sensitive to hardware noise as two-qubit gate counts increase.

In practice, the preferred formulation depends on the dominant hardware constraint:
\begin{itemize}
    \item If the primary bottleneck is \emph{available qubits}, HUBO may be attractive despite deeper circuits, particularly if compilation and error mitigation for multi-qubit interactions continue to improve.
    \item If the primary bottleneck is \emph{(two-qubit) gate fidelity}, QUBO may remain preferable because it restricts cost terms to one- and two-body interactions.
\end{itemize}
These trade-offs will evolve with hardware improvements. 
In the near term, limited coherence times and two-qubit gate errors strongly penalise the deeper circuits induced by higher-order cost terms, favouring QUBO-style formulations when qubit counts are manageable.
As fidelities and connectivity improve, the relative cost of implementing multi-qubit interactions should decrease: deeper compiled circuits will be less noise-dominated, and improvements in connectivity and routing will reduce SWAP overheads that currently inflate depth. 
In such a regime, qubit count rather than circuit depth may become the dominant constraint, increasing the attractiveness of higher-order encodings. More generally, this suggests a transition from a depth-limited regime to a memory-limited regime as hardware matures.
\Cref{sec:noise_sensitivity} analyses the impact of hardware fidelity on sampling requirements in greater detail.

\subsection{Role of the Iterative-QAOA heuristic}
Iterative-QAOA combines a fixed linear-ramp schedule (LR-QAOA) with iterative updates of initial-state biases, steering the sampling distribution toward low-energy regions. 
This approach avoids the overhead of full variational optimisation while achieving substantially stronger concentration on low-energy states than standard shallow-depth QAOA.

In our experiments, this update rule can compensate for shallow depth in some regimes (notably smaller QUBO and HUBO instances), but it can also amplify biases toward suboptimal basins when the schedule is mismatched to the instance or when the number of effective samples is small. 
This motivates future work on: (i) schedule selection that is robust across depths and instance families, (ii) principled stopping criteria and restarts when convergence to a suboptimal mode is detected, and (iii) adaptive choices of CVaR fraction $\alpha$ as a function of noise level and iteration.

\subsection{Limitations and future directions}
This study is a proof-of-principle exploration rather than an end-to-end assembly evaluation, and several limitations point to clear next steps. 
First, our benchmarks focus on synthetic instances of the Oriented Tangle Resolution subproblem, and the relationship between instance structure and Iterative-QAOA performance warrants a more systematic study. 
Second, while CVaR post-selection improves robustness under noise, it increases total shot requirements; efficient allocation of finite sampling budgets will be important for practical workflows.

Finally, it remains necessary to connect optimisation performance to downstream PGSA metrics, including contiguity, correctness, and robustness to mapping ambiguity and copy-number error. Integrating these optimisation methods into the full PGSA workflow described in~\cite{cudbyPangenomeguidedSequenceAssembly2026} will allow direct evaluation of assembly-level impact. Demonstrating that improvements in the Oriented Tangle Resolution objective translate into measurable gains over specialised classical solvers will be essential for assessing practical relevance.

%% file: input_files_v1/methods.tex
\subsection{Problem formulation}
We consider two binary optimisation formulations of~\cref{prob:oriented_tangle}.

The first is a QUBO, in which all terms have order 2; equivalently, the objective is to minimise
\begin{equation}
     Q(x) \coloneqq x^TMx,
\end{equation}
where $M\in \mathbb{R}^{n \times n}$ is a symmetric, real-valued matrix.
QUBO encodings are expressive and capture classical problems such as MAX-CUT and the Travelling Salesman Problem (TSP). 
However, graph-traversal QUBO encodings require $\mathcal{O}(N^2)$ variables in the graph size $N$, because they enumerate node–time pairs. 
The QUBO we use here is a slight modification of the formulation in~\cite{cudbyPangenomeguidedSequenceAssembly2026} and requires $\mathcal{O}(N^2)$ binary variables for an $N$-node graph.

The second is a HUBO, where the objective is a polynomial $H:\{0,1\}^n\to\mathbb{R}$ with terms of arbitrary degree. 
By encoding node indices in binary, the HUBO requires only $\mathcal{O}(N\log N)$ variables for an $N$-node graph. 
In classical practice, higher-order terms are often reduced to quadratic form via ancilla variables; here we instead encode the higher-order form directly to prioritise qubit efficiency.
\subsubsection{Quadratic unconstrained binary optimisation}
We use binary variables $x_{t,v,b}\in\{0,1\}$, where $x_{t,v,b}=1$ indicates that the path visits vertex $v$ with orientation $b\in\{0,1\}$ at time $t$. The index $v$ ranges over the graph vertex set $V$ and $t\in\{1,\ldots,T\}$ indexes the walk positions.

We initially select the walk length $T = \sum_{v \in V_+} w(v)$, i.e., the total desired visits inferred from positive-orientation copy-number estimates. 
If required, $T$ may be adjusted up or down after observing preliminary solutions; in practice, this search is rarely necessary to obtain high-quality results.

The QUBO cost function has three components: a one-hot constraint per time step, an edge-following constraint between successive times, and a node-frequency matching term. Writing $Q_G(\{x_{t,v,b}\}) = Q^1_G + Q^2_G + Q^3_G$, we define
\begin{widetext}
\begin{align}
    Q^1_G(\{x_{t,v,b}\}) \coloneqq \Lambda_1&
        \sum_{t=1}^T \left( \sum_{v \in V}\sum_{b \in \{0,\,1\}}x_{t,v,b} -1\right)^2,
    \\
    Q^2_G(\{x_{t,v,b}\}) \coloneqq \Lambda_2 &
        \sum_{t=1}^{T-1}\left(
        1 - \sum_{v,\,v' \in V}\sum_{b,\,b' \in \{0,\,1\}}  x_{t,v,b}x_{t+1,v',b} 
        \mathbbm{1}\Bigl(
        \bigl((v,\,b),(v',\,b')\bigr) \in E
        \Bigr) 
        \right),
    \\
    Q^3_G(\{x_{t,v,b}\}) \coloneqq  \phantom{\Lambda_3}&
    \sum_{v \in V} \left( \sum_{t=1}^T x_{t,v,0} + x_{t,v,1} - w(v) \right)^2.
\end{align}
\end{widetext}
The Lagrange multipliers $\Lambda_1,\Lambda_2$ must be chosen large enough that feasible (constraint-satisfying) assignments have lower objective than infeasible ones, but not so large that the constrained subspace forms steep, ill-conditioned valleys that overwhelm the optimiser. 
Empirically, intermediate values such as $\Lambda_1=10$ and $\Lambda_2=5$ perform well on our instances.

\subsubsection{Higher-order unconstrained binary optimisation}
Let $n=\lceil\log_2(2N)\rceil$ and introduce binary variables $x_{t,k}$ for $t=1,\ldots,T$ and $k=0,\ldots,n-1$.
For each time $t$, the bits $x_{t,0},\ldots,x_{t,n-1}$ encode an integer $X_t\in\{0,\ldots,2^n-1\}$.
We associate vertex indices $0,\ldots,2N-1$ to $V=V_+\cup V_-$, using even integers for positively oriented vertices and odd integers for negatively oriented ones. 
If every $X_t<2N$, then $(X_1,\ldots,X_T)$ corresponds to a walk; otherwise the assignment is infeasible.

The HUBO formulation relies heavily on the following identity for binary variables $x$ and $y$:
\begin{equation}
\label{eq:binary_indicator}
    1 - x - y + 2xy = \begin{cases}
        1 & x = y,\\
        0 & x \neq y.
    \end{cases}
\end{equation}
By taking the product of $n$ terms of the form of the LHS of~\cref{eq:binary_indicator}, we can construct an indicator function for the encoded integers $X_t$.
For an integer $i \in \{0,\ldots,2N\}$, let $(b_0,\,b_1,\ldots,b_{n-1})$ be its binary encoding.
Then we have
\begin{equation}
    \mathbbm{1}(X_t = i) \coloneqq  \prod_{k=0}^{n-1} \Bigl(1 - b_k - x_{t,k} + 2b_k \cdot x_{t,k}\Bigr) = \begin{cases}
        1 & X_t = i,\\
        0 & \text{else.}
    \end{cases}
\end{equation}

The HUBO cost comprises an edge-penalty term and a node-frequency term: $H_G(\{x\}) = H^1_G + H^2_G$, where
\begin{widetext}
\begin{align}
        H^1_G(\{x_{t,\,k}\}) \coloneqq & \Lambda_1 \sum_{t=1}^{T-1} \Biggl[ 
        1
        -\sum_{i =0}^{2N-1} \biggl(
            \mathbbm{1}(X_t = i)
            \cdot
                \sum_{{j:(i,j) \in E}} \mathbbm{1}(X_{t+1} = j)
            \biggr)
        \Biggr],
    \\
    H^2_G(\{x_{t,\,k}\}) \coloneqq & \phantom{\Lambda_2} \sum_{l =0}^{N-1} \Biggl[ 
        \sum_{t=1}^T  
        \Bigl(
            \mathbbm{1}(X_t = 2l) + \mathbbm{1}(X_t = 2l + 1) 
        \Bigr)
        - w(2l)
    \Biggr]^2.
\end{align}
\end{widetext}

\subsubsection{Mapping binary optimisation to quantum optimisation}
\label{subsec:bin_to_quantum}
We convert binary cost functions to quantum Hamiltonians by substituting each bit $x_i$ with the operator $\tfrac{1}{2}(I-Z_i)$. 
For a cost function $C(\{x_i\})$ this yields a Hamiltonian $H_C$ whose computational-basis energies satisfy $\bra{x}H_C\ket{x}=C(x)$.

In QAOA the cost-layer unitary is $\exp(-i\gamma H_C)$. 
Since all $Z$ operators commute, this exponential decomposes into a product of commuting $Z$-interaction terms.
The maximum number of qubits appearing in any interaction equals the degree of the cost polynomial; in particular, QUBO cost functions involve only one- and two-qubit $Z$ interactions.


\subsection{Quantum circuit compilation}
\label{subsec:compilation}
On contemporary hardware, minimising circuit depth and two-qubit gate count is essential to reduce noise. 
In QAOA the initialisation and mixer layers are typically single-qubit layers; the dominant compilation challenge is therefore the cost-layer exponentiation $\exp(-i\gamma H_C)$.
Since this operator is given by a product of commuting $Z$ rotations, it is usually possible to find representations that are far more compact than a naive decomposition would suggest.

\begin{figure}
    \centering
    \includegraphics[width=\columnwidth]{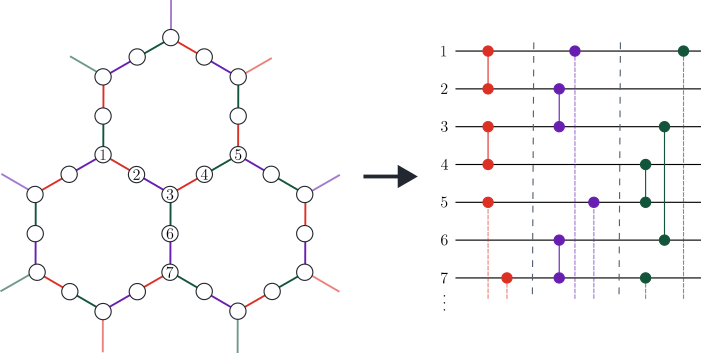}
    \caption[
    Heavy-hex edge colouring
    ]{
    The use of edge colouring to efficiently implement two-qubit gates on the IBM heavy-hex layout. 
    (Left) Three cells of the heavy-hex layout and the corresponding 3-colouring. Some of the qubits are assigned a virtual qubit index.
    (Right) Part of a quantum circuit that implements all pairwise interactions of a heavy-hex layout in circuit depth 3. 
    Vertical barriers separate operations applied at different circuit layers.
    }
    \label{fig:heavy_hex_colouring}
\end{figure}

For QUBO-QAOA on limited-connectivity layouts (e.g., heavy-hex), existing SWAP strategies and layout algorithms yield compact implementations~\cite{weidenfeller_scaling_2022,matsuo_sat_2023}. 
Typical approaches (i) compute a qubit layout that minimises required SWAP depth and (ii) order two-qubit interactions using an edge-colouring of the hardware coupling graph so that interactions in each time slice can be applied in parallel. 
For heavy-hex connectivity, a 3-colouring suffices to implement any set of pairwise interactions between two SWAP layers in at most depth~3.
An illustrative example is provided in~\cref{fig:heavy_hex_colouring}.

For HUBO-QAOA, additional complications arise because cost terms involve multi-qubit $Z$ rotations.
We extend the SWAP and layout strategies to increase the number of higher-order interactions that become local after a minimal number of SWAP layers, and we generalise the CNF-based layout search of Matsuo et al.~\cite{matsuo_sat_2023} to account for higher-order terms. 
Because the CNF can become large, we cap the maximum interaction size considered in the formula (we typically use size $\leq 6$) and then solve a MAX-SAT instance to find a layout that implements as many interactions as possible for a chosen SWAP depth $d$. 
Remaining interactions are implemented manually using SWAPs. 
We select $d$ by sampling candidate depths and choosing the one minimising overall compiled depth, balancing the trade-off between SWAP layers and leftover manual implementations.
We use the MAXSAT solver \textit{NuWLS}~\cite{chuNuWLSImprovingLocal2023}.

\begin{figure}[t]
    \centering
    \includegraphics[width=\columnwidth]{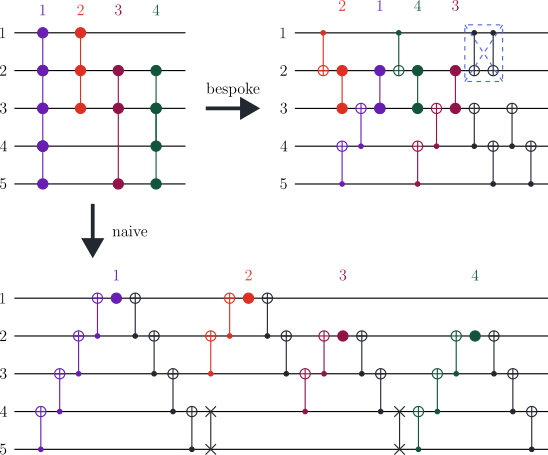}
    \caption[
    Bespoke HUBO-QAOA transpilation
    ]{
    An example of bespoke HUBO-QAOA transpilation on a linear nearest-neighbour topology.
    Dots represent qubits participating in a $Z$-rotation.
    (Upper-left) A sequence of 4 interactions involving 3 to 5 qubits. 
    (Upper-right) The compiled circuit, using our transpilation strategy, comprises 10 CX gates and 4 $R_{zz}$ gates. The orders of interactions 1 \& 2 and 3 \& 4 were swapped. Interaction 3 was implemented without any SWAP gates, even though it was not connected in the topology. Two of the CX gates added as part of the transpilation logic can be trivially cancelled. 
    (Lower) The naively compiled circuit comprises 22 CX gates, 2 SWAP gates and 4 $R_z$ gates. \textit{Qiskit}'s inbuilt transpiler at the highest optimisation level achieves only two fewer two-qubit gates.
    }
    \label{fig:interaction_chain}
\end{figure}

To compile multi-qubit $Z$-rotations efficiently, we avoid the naive CX ladder implementation where possible. 
We implemented a bespoke strategy which performs multi-qubit rotations via a $R_{zz}$ on a pair of target qubits, with CX networks to collect parity. 
We search for sequences of interactions that can be executed in a fixed qubit location, reusing partial parity information to cancel CXs across successive rotations. 
This approach yields substantial CX cancellations compared to naive transpilation; see \cref{fig:interaction_chain} for an illustrative example on a linear nearest-neighbour layout.

In~\cref{tab:circuit_depth}, we compare the results of using our bespoke transpiler with the inbuilt \textit{Qiskit} transpiler at the highest level of optimisation.
We achieve at least a $43\%$ and up to a $67\%$ reduction in circuit depth compared to the \textit{Qiskit} transpiler, which already contains advanced circuit compilation and layout techniques.

\begin{table}[t]
\caption{
A table comparing the two-qubit gate count and two-qubit depth of HUBO-QAOA cost Hamiltonians on a grid topology across different circuit transpilation methods. 
The ``\textit{Qiskit}'' columns refer to using the standard \textit{Qiskit} transpiler with optimisation level 3. 
The ``Custom'' columns refer to the use of our custom transpiler.
}
\centering
\renewcommand{\arraystretch}{1.2}
\begin{tabular}
{
@{} 
l
*{3}{r} @{\hskip 5\tabcolsep}
*{3}{r}
@{}
}
\toprule\toprule
\multirow{2}{*}{Qubits} 
& \multicolumn{2}{@{}c@{\hskip 3\tabcolsep}}{$2Q$ count} 
& \multirow{2}{*}{\begin{tabular}[t]{@{}r@{}}\% count\\reduction\end{tabular}}
& \multicolumn{2}{@{}c@{\hskip 3\tabcolsep}}{$2Q$ depth} 
& \multirow{2}{*}{\begin{tabular}[t]{@{}r@{}}\% depth\\reduction\end{tabular}}
\\ \cmidrule(l{\dimexpr \tabcolsep}r{\dimexpr \tabcolsep}){2-3} 
\cmidrule(r{\dimexpr \tabcolsep}){5-6} 
 & \textit{Qiskit} & Custom & & \textit{Qiskit} & Custom &  \\ \midrule
 4 & 21 & 10 & 52.4 & 18 & 6 & 66.7 \\ 
 8 & 823 & 522 & 36.6 & 743 & 410 & 44.8 \\ 
 9 & 365 & 182 & 50.1 & 322 & 138 & 57.1 \\ 
 8 & 824 & 543 & 34.1 & 741 & 424 & 42.8 \\ 
 12 & 605 & 351 & 42.0 & 513 & 179 & 65.1 \\ 
 12 & 1732 & 1222 & 29.4 & 1522 & 751 & 50.7 \\ 
  15 & 817 & 630 & 22.9 & 705 & 306 & 56.6 \\ 
 16 & 3108 & 1935 & 37.7 & 2612 & 1049 & 59.8 \\ 
  \bottomrule\bottomrule
\end{tabular}
\label{tab:circuit_depth}
\end{table}
\renewcommand{\arraystretch}{1}

\subsection{Non-variational quantum optimisation}
\label{subsec:iterative_qaoa}
We use an Iterative-QAOA procedure introduced in~\cite{lopez-ruiz_non-variational_2025} 
that combines ideas from warm-start QAOA~\cite{egger_warm-starting_2021,WILLSCH2022108411,grant_initialization_2019} 
with a linear-ramp (LR) parameter schedule~\cite{montanez-barrera_toward_2025,dehn_extrapolation_2025,kremenetski_quantum_2023} 
Iterative-QAOA updates initial-state biases each iteration using measurement data from the previous run, avoiding a full variational optimisation loop.

A depth-$p$ QAOA circuit prepares the state
\begin{equation}
      \prod_{k=p}^1 \Bigl[\exp(-i\beta_k H_M)\exp(-i\gamma_k H_C)\Bigr]\, O \ket{0}^{\otimes q}.
\end{equation}
Following LR-QAOA we set
\begin{equation}
    \beta_k \coloneqq  \Bigl(1 - \frac{2k-1}{2p}\Bigr) \Delta_\beta, \qquad
    \gamma_k \coloneqq  \frac{2k-1}{2p} \Delta_\gamma,
\end{equation}
with hyperparameters $\Delta_\beta,\Delta_\gamma$ chosen by empirical search; see \cref{sec:param_exploration} for details.
In our experiments we used $\Delta_\beta=0.63,\ \Delta_\gamma=0.16$ for QUBO and $\Delta_\beta=0.75,\ \Delta_\gamma=0.30$ for HUBO.

The initial state at iteration $j$ is a product state with independent bit-flip probabilities:
\begin{equation}
    \ket{\psi^{(j)}_{\mathrm{init}}} = \bigotimes_{i=1}^q \Bigl(\sqrt{1-p_i^{(j)}}\ket{0} + \sqrt{p_i^{(j)}}\ket{1}\Bigr),
\end{equation}
implemented by single-layer $R_y$ rotations with angles $\phi_i^{(j)} = 2\arcsin\sqrt{p_i^{(j)}}$. 
The mixer Hamiltonian is chosen so that $\ket{\psi^{(j)}_{\mathrm{init}}}$ is an eigenstate:
\begin{equation}
    H_M^{(j)} = \sum_{i=1}^q \bigl(-\sin(\phi_i^{(j)}) X_i - \cos(\phi_i^{(j)}) Z_i \bigr).
\end{equation}
The rotation $e^{-i\beta H_M^{(j)}}$ can be implemented in depth~3 by applying $R_y(\phi_i^{(j)})\,R_z(-2\beta)\,R_y(-\phi_i^{(j)})$ to each qubit.

After measuring a batch of outcomes $\{s_k^{(j)}\}$, with energies $E_k^{(j)}=\bra{s_k^{(j)}}H_C\ket{s_k^{(j)}}$, we form a weighted distribution
\begin{equation}
    P^{(j)}(s_k^{(j)}) \;=\; \frac{\exp\bigl(-\beta_T^{(j)} (E_k^{(j)})^2\bigr)}{\sum_{k'}\exp\bigl(-\beta_T^{(j)} (E_{k'}^{(j)})^2\bigr)},
\end{equation}
where $\beta_T^{(j)}$ is a feedback inverse-temperature parameter controlling feedback strength. 
The $Z$-expectation under this weighting is
\begin{equation}
    \langle Z_i^{(j)}\rangle \; \coloneqq \; \sum_k P^{(j)}(s_k^{(j)}) \bra{s_k^{(j)}}Z_i\ket{s_k^{(j)}}.
\end{equation}

To reduce the influence of noisy, high-energy outcomes, we optionally apply a CVaR-style filter: keep only the best-performing fraction $\alpha$ of samples (by energy) when computing weighted expectations~\cite{barkoutsos_improving_2020,barron_provable_2024}.
Based on empirical performance, we use $\alpha\in[0.05,0.1]$.  

We then update probabilities as
\begin{equation}
    p_i^{(j+1)} = \tfrac{1}{2}\bigl(1 - \langle Z_i^{(j)}\rangle\bigr).
\end{equation}
To prevent the prior from becoming overly concentrated and to maintain exploration, we clip probabilities to $[\epsilon,1-\epsilon]$ with $\epsilon=0.15$:
\[
p_i^{(j+1)} \leftarrow \min\bigl(1-\epsilon,\max(\epsilon,\,p_i^{(j+1)})\bigr).
\]
In practice, we run up to five iterations and use a quadratic schedule for $\beta_T^{(j)}$ with $\beta_T^{(1)}=0.015$ and $\beta_T^{(5)}=0.045$.

For QUBO problems, variables at each time $t$ form a one-hot encoding, so we initialise with $p_i^{(0)} = 1/(2N)$ to bias the prior toward valid walks. For HUBO problems, no prior structure is assumed, and we initialise with unbiased $p_i^{(0)}=1/2$.

%% file: input_files_v1/appendix.tex
\section{A primer on genomics and sequence assembly}
\label{sec:bio_primer}

This appendix briefly introduces the biological and computational concepts underlying pangenome-guided sequence assembly (PGSA).

A \emph{genome} is the full DNA sequence of an individual and can be modelled as a string over the four-letter alphabet $\Sigma=\{A,C,G,T\}$. Concretely, $\mathcal{G}\in\Sigma^L$ with $L$ large (for humans $L\approx 3\times10^9$). DNA is double-stranded: each strand has a \emph{reverse-complement}, obtained by reversing the sequence and substituting $A\leftrightarrow T$, $C\leftrightarrow G$.

Sequencing technologies produce many short, noisy fragments called \emph{reads} rather than the full genome. Formally, an experiment yields a multiset $\{r_1,\dots,r_m\}$ where each $r_i$ is a substring of $\mathcal{G}$ observed with errors. Experiments are typically run to sufficient depth (e.g., $\sim30\times$ coverage) so that most genomic positions are sampled multiple times; this redundancy is essential for error correction and disambiguation.

\emph{De novo} sequence assembly asks:

\begin{problem}
    Given a multiset of noisy substrings sampled from an unknown string $\mathcal{G}$, reconstruct $\mathcal{G}$.
\end{problem}

In idealised, noiseless settings this reduces to shortest-superstring or Eulerian-path formulations. In practice, sequencing errors, repeats, and structural variation complicate the problem and motivate heuristic and optimisation approaches.

A common practical strategy is reference-guided assembly: align reads to a curated representative genome (e.g., GRCh38~\cite{schneider_evaluation_2017}), group reads by location, and infer local consensus. Reference-guided methods are efficient and effective for correcting small errors but can introduce \emph{reference bias}: regions that differ substantially from the reference may be misassembled or missed.

\emph{Pangenomes} mitigate reference bias by representing population-level variation in a graph structure rather than a single linear sequence. A pangenome encodes alternative subsequences and structural variants within a unified graph. Under this model, an individual's genome corresponds approximately to a path (or walk) through the pangenome graph. Reads are mapped to graph nodes or edges, producing counts or weights that indicate the presence and approximate frequency of graph components.

This observation underlies the PGSA workflow: infer a walk through the pangenome graph that best explains the read-derived weights while respecting path-consistency constraints. In our formulation this becomes a structured binary optimisation problem where decision variables encode component inclusion and penalty terms enforce path constraints.

\section{LR-QAOA parameter exploration}
\label{sec:param_exploration}
\begin{figure*}[t]
    \centering
    \begin{subfigure}[c]{0.49\textwidth}
        \includegraphics[width=\textwidth]{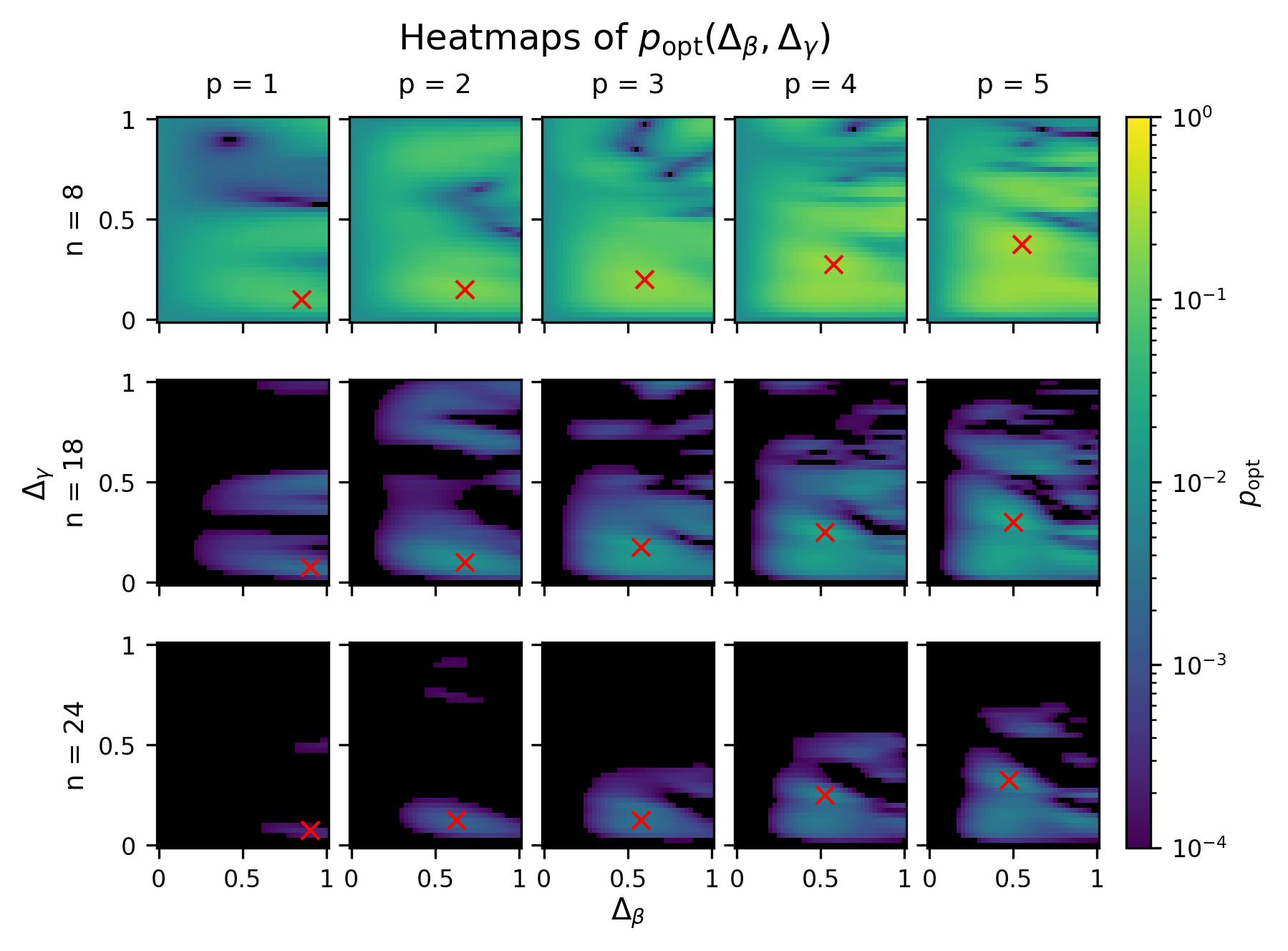}
        \caption{
        QUBO-LR-QAOA heatmaps. 
        We choose $\Delta_\beta = 0.63$, $\Delta_\gamma = 0.16$ as they perform well across instances at low $p$.
        }
        \label{fig:qubo_linear_param_exploration}
    \end{subfigure}
    \hfill
    \begin{subfigure}[c]{0.49\textwidth}
        \includegraphics[width=\textwidth]{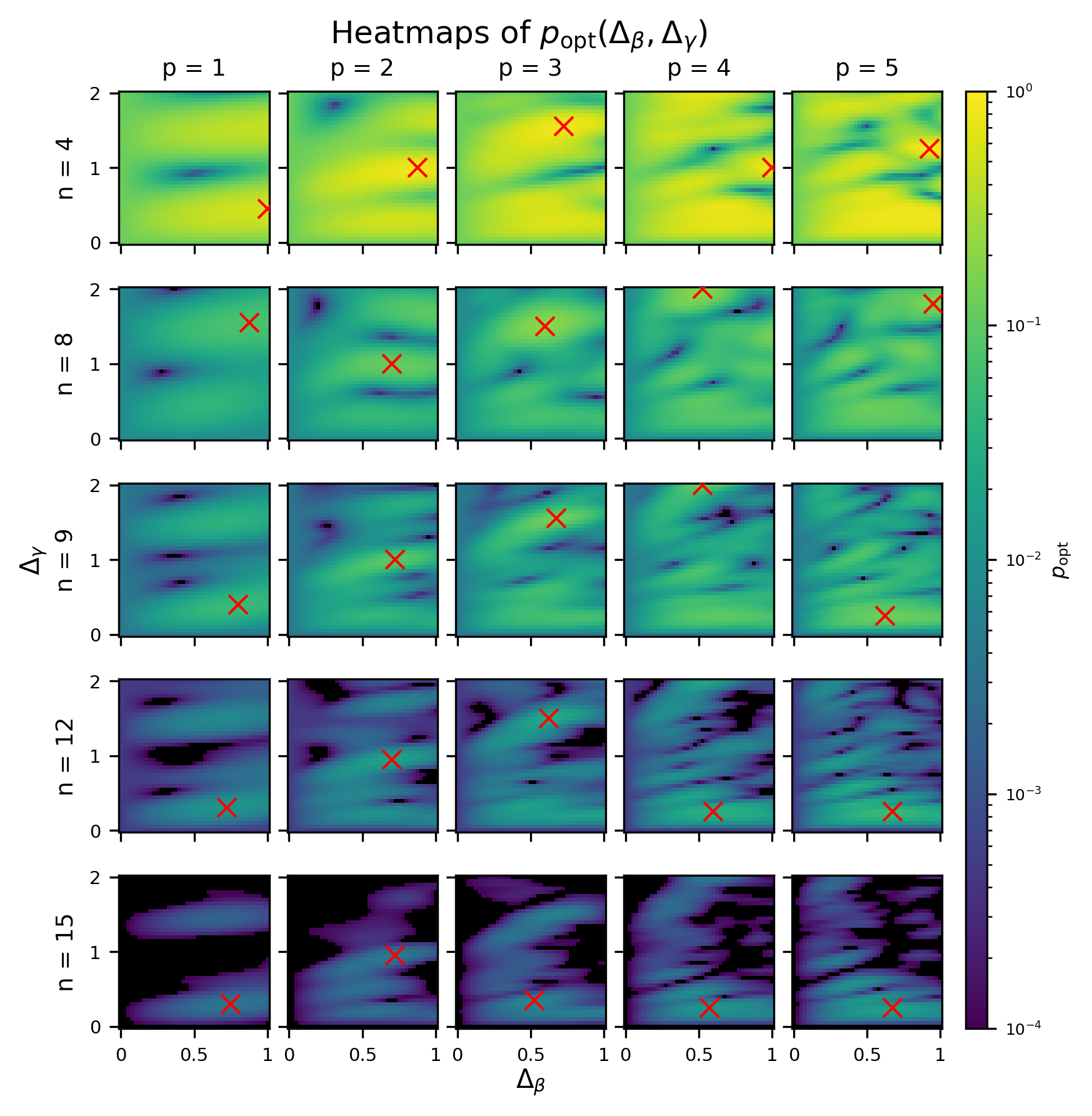}
        \caption{
        HUBO-LR-QAOA heatmaps. 
        We choose $\Delta_\beta = 0.75$, $\Delta_\gamma = 0.30$ as they perform well across instances at $p=1$.
        }
        \label{fig:hubo_linear_param_exploration}
    \end{subfigure}
    \caption{    
    Heatmaps of the probability of sampling an optimal solution, $p_{\text{opt}}$, for LR-QAOA over a grid of schedule hyperparameters $(\Delta_{\beta},\,\Delta_{\gamma})$. Each panel corresponds to a different small problem instance and circuit depth $p$, illustrating both the strong dependence of performance on the schedule parameters and the extent to which well-performing regions persist (or shift) as $p$ increases.
    }
    \label{fig:param_exploration}
\end{figure*}

LR-QAOA performance depends strongly on the schedule hyperparameters $(\Delta_\beta,\Delta_\gamma)$ and on circuit depth $p$. 
We measure performance by the probability of sampling an optimal solution, $p_{\text{opt}}$, using noiseless statevector simulation. 
For each instance and each $p$ we sweep a fine grid of $(\Delta_\beta,\Delta_\gamma)$ to build heatmaps of $p_{\text{opt}}$; representative results are shown in~\cref{fig:qubo_linear_param_exploration} (QUBO) and~\cref{fig:hubo_linear_param_exploration} (HUBO).

Across QUBO instances, we observe a consistent trend: at $p=1$ the best schedules favour relatively large $\Delta_\beta$ and small $\Delta_\gamma$, and as $p$ increases the locally optimal point drifts approximately along a smooth trajectory toward intermediate values of both parameters. 
We therefore selected $(\Delta_\beta,\Delta_\gamma)=(0.63,0.16)$ as a single fixed schedule that performs robustly across the small-$p$ regime explored in our experiments.

HUBO instances are more heterogeneous, but at $p=1$ the pair $(\Delta_\beta,\Delta_\gamma)=(0.75,0.30)$ performs well across our representative cases, so we adopt this fixed schedule for HUBO benchmarks. 
In general, we observe that HUBO performance at small depth is often less sensitive to $\Delta_\beta$ across a broad range when $\Delta_\gamma$ is small.

\begin{figure*}[t]
    \centering
    \begin{subfigure}[c]{0.49\textwidth}
        \includegraphics[width=\textwidth]{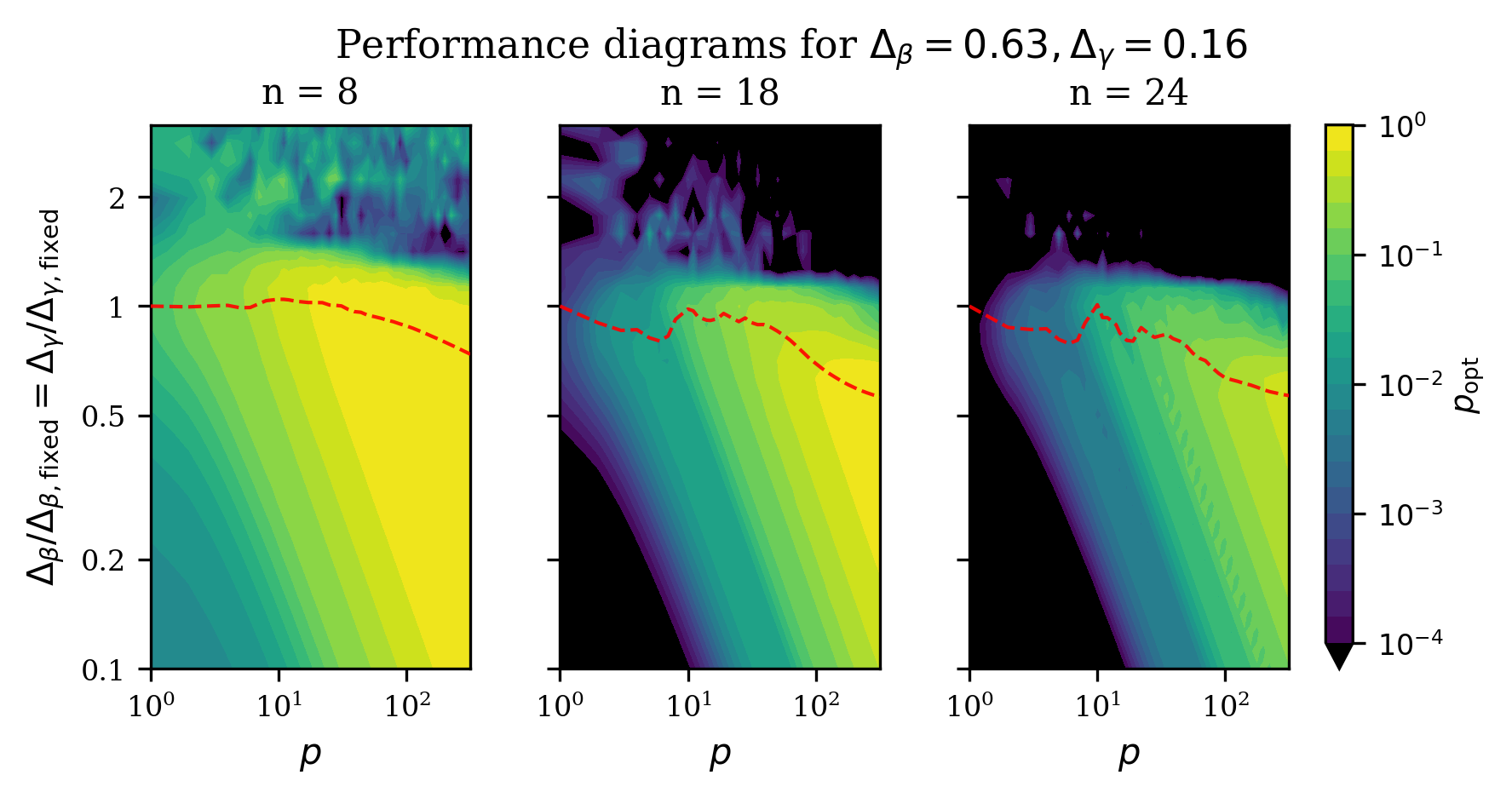}
        \caption{
        QUBO-LR-QAOA performance diagrams. 
        }
        \label{fig:qubo_performance_diagrams}
    \end{subfigure}
    \hfill
    \begin{subfigure}[c]{0.49\textwidth}
        \includegraphics[width=\textwidth]{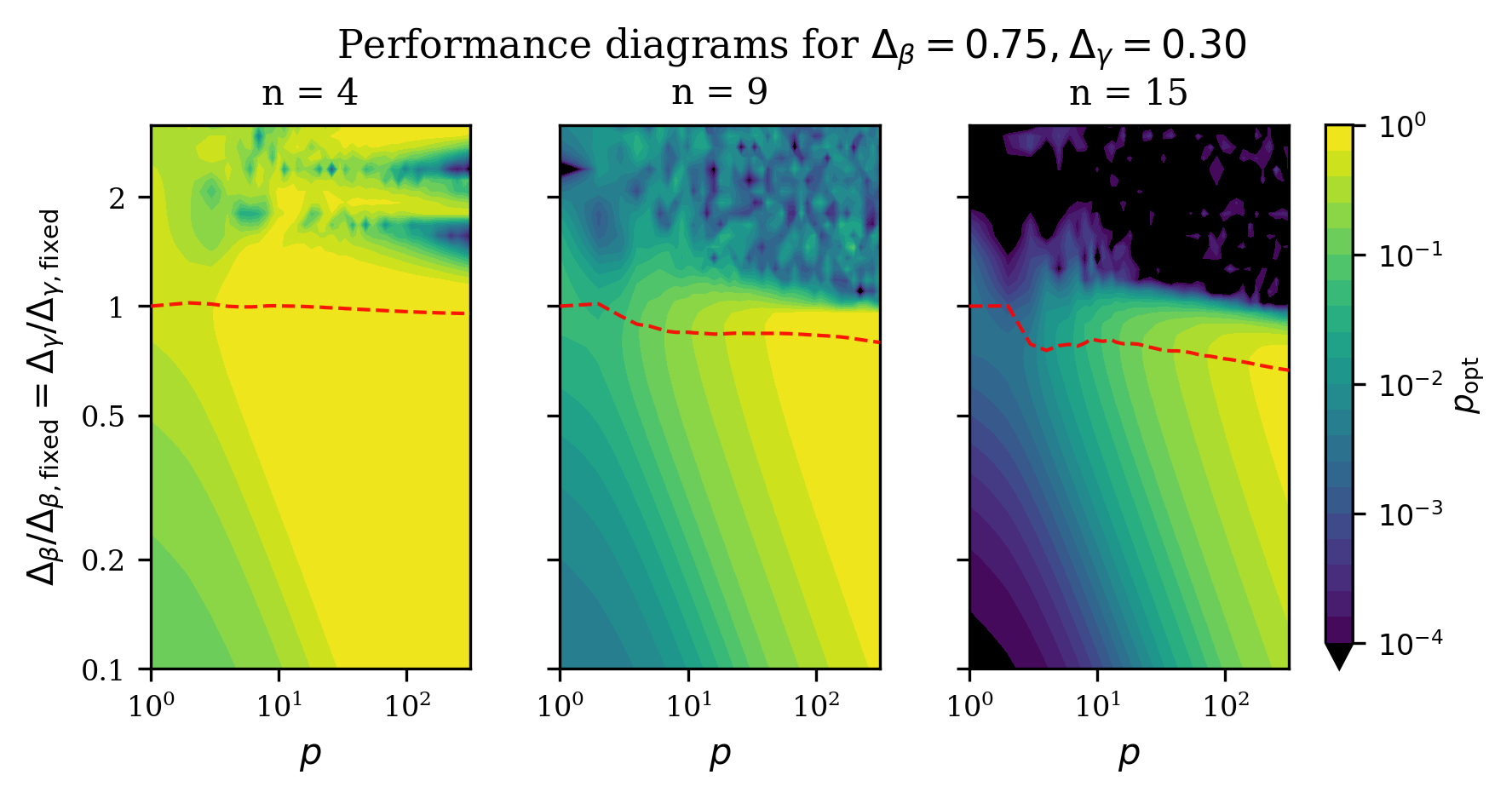}
        \caption{
        HUBO-LR-QAOA performance diagrams. 
        }
        \label{fig:hubo_performance_diagrams}
    \end{subfigure}
    \caption{    
    Performance diagrams for LR-QAOA, showing the dependence of the probability of sampling an optimal solution, $p_{\text{opt}}$, on circuit depth $p$ and on schedule hyperparameters $(\Delta_\beta,\,\Delta_\gamma)$ along rays of fixed ratio $\Delta_\beta/\Delta_\gamma$.
    The horizontal axis corresponds to increasing depth $p$.
    The vertical axis parameterises a joint rescaling of $(\Delta_\beta,\,\Delta_\gamma)$ with $\Delta_\beta/\Delta_\gamma$ held fixed, so that moving vertically explores schedules with the same ramp ``shape'' but different overall range.
    Each panel corresponds to a different HUBO instance, and colour indicates $p_{\text{opt}}$ obtained from noiseless simulations.
    The red dashed curve indicates a gradient-ascent path in this parameter space starting from the fixed shallow-depth choice $(\Delta_\beta,\,\Delta_\gamma)=(0.75,\,0.30)$ at $p=1$, illustrating how the locally optimal schedule drifts as $p$ increases.
    }
    \label{fig:performance_diagrams}
\end{figure*}

To obtain a clearer picture of the dependence on parameter schedules, we also prepare performance diagrams~\cite{kremenetski_quantum_2021,kremenetski_quantum_2023} across the parameter space for both QUBO and HUBO problems, which capture the algorithm’s performance over different parameter values and circuit depths.
These are shown in~\cref{fig:qubo_performance_diagrams,fig:hubo_performance_diagrams}.
In these figures, the x-axis corresponds to increasing circuit depth $p$, and the y-axis represents different rescalings of $\Delta_\beta$ and $\Delta_\gamma$ with the ratio $\Delta_\beta / \Delta_\gamma$ held fixed.
The red dashed lines show a gradient ascent up the ridge of increasing $p_\text{opt}$ starting from $p=1$, $\Delta_{\beta/\gamma} = \Delta_{\beta/\gamma,\,\text{fixed}}$.

To visualise schedule dependence as depth varies, we construct performance diagrams~\cite{kremenetski_quantum_2021,kremenetski_quantum_2023} (see \cref{fig:qubo_performance_diagrams,fig:hubo_performance_diagrams}). 
Each diagram shows $p_{\text{opt}}$ as a function of depth $p$ and a joint rescaling of $(\Delta_\beta,\Delta_\gamma)$ along rays of fixed ratio $\Delta_\beta/\Delta_\gamma$. 
The diagrams confirm the shallow-depth robustness of our chosen schedules and illustrate the predictable drift of locally optimal schedules as $p$ increases.

\section{Sensitivity to hardware noise}
\label{sec:noise_sensitivity}
The effectiveness of CVaR-style post-selection depends on the probability that a circuit execution yields a sample sufficiently low in energy to be useful for the bias update.
For circuits with thousands of two-qubit gates, modest changes in per-gate error rates translate into large changes in the oversampling required to obtain a fixed number of high-quality samples.

As an illustrative estimate, consider the \textit{IBM Boston} device. Its reported median two-qubit error rate is $e_{\mathrm{med}}=1.24\times10^{-3}$, and its layered two-qubit error rate for a long chain is $e_{\mathrm{layer}}=2.10\times10^{-3}$. 
For a representative 48-qubit QUBO circuit with $G=2{,}865$ two-qubit gates, a simple independent-error approximation gives the per-shot probability of experiencing zero two-qubit errors as approximately $(1-e)^G$. 
Using $e_{\mathrm{med}}$ yields
$p_{\mathrm{good}}\approx(1-e_{\mathrm{med}})^{2865}\approx 2.9\%,$
whereas using $e_{\mathrm{layer}}$ gives $p_{\mathrm{good}}\approx 0.24\%$.
To collect on the order of $M\!=\!4{,}000$ good samples, one therefore needs roughly $M/p_{\mathrm{good}}$ total shots, i.e., on the order of $1.4\times10^5$ shots under the median rate and $\sim1.6\times10^6$ shots under the layered rate.

This back-of-envelope calculation omits correlated errors and the fact that imperfect shots may still contribute information, but it highlights a practical point: the CVaR fraction $\alpha$ and the required oversampling scale inversely with the per-shot good probability, which itself is highly sensitive to two-qubit error rates. 
Consequently, modest improvements in two-qubit fidelities or reductions in effective layered error can dramatically reduce sampling requirements, expanding the accessible regime of circuit depth and problem size for Iterative-QAOA methods.